\documentclass{emulateapj}
\shorttitle{Twelve years of spectroscopic monitoring in the Galactic}
\shortauthors{Habibi et al.}

\begin{document}

\title{Twelve years of spectroscopic monitoring in the Galactic Center: the closest look at S-stars near the black hole}

\author{M.~Habibi\altaffilmark{1}, S.~Gillessen\altaffilmark{1}, F.~Martins\altaffilmark{2}, F.~Eisenhauer\altaffilmark{1}, P. M.~Plewa\altaffilmark{1}, O.~Pfuhl\altaffilmark{1},E.~ George\altaffilmark{1}, J.~Dexter\altaffilmark{1}, I.~ Waisberg\altaffilmark{1}, T.~Ott\altaffilmark{1}, S.~von~Fellenberg\altaffilmark{1}, M.~ Baub\"ock, A.~ Jimenez-Rosales, and R.~Genzel\altaffilmark{1,3}}
\altaffiltext{1}{Max-Planck-Institut f\"ur Extraterrestrische Physik, 85748 Garching, Germany}
\altaffiltext{2}{LUPM, Universit\'e de Montpellier, CNRS, Place Eug\`ene Bataillon, F-34095 Montpellier, France}
\altaffiltext{3}{Physics Department, University of California, Berkeley, CA 94720, USA}

\begin{abstract}
We study the young S-stars within  a distance of 0.04 pc from the supermassive black hole in the center of our Galaxy.  Given how inhospitable the region is for star formation, their  presence is more puzzling the younger we estimate their ages. In this study, we analyse the result of 12 years of high resolution spectroscopy within the central arcsecond of the Galactic Center (GC). By co-adding between 55 and 105 hours of spectra we have obtained high signal to noise H- and K-band spectra of eight stars orbiting the central supermassive black hole. Using deep H-band spectra, we show that these stars must be high surface gravity (dwarf) stars. We compare these deep spectra to detailed model atmospheres and stellar evolution models  to infer the stellar parameters. Our analysis  reveals an effective temperature of 21000--28500 K, a rotational velocity of 60-170 km/s, and a surface gravity of 4.1--4.2. These parameters imply a spectral type of B0-B3V for these stars. The inferred masses lie within 8--14$M_\odot$. We derive an age of  $6.6^{+3.4 }_{-4.7}$ Myr for the star S2, which is compatible with the age of  the clockwise rotating  young stellar disk in the GC. We estimate the age of all other studied S-stars to be less than 15 Myr, which are compatible with the age of S2 within the uncertainties. The relatively low ages for these S-stars favor a scenario in which the stars formed in a local disk rather than the field-binary-disruption scenario  throughout a longer period of time.

\end{abstract}

\section{Introduction}
\label{sec:intro}
Infrared observations of  the central parsec of our Galaxy have revealed a  population of nearly 200 young, massive stars  \citep[e.g.][]{genz94, blum95, Bartk10, Do2013}. The existence of young stars in the extreme environment so close to a supermassive black hole (SMBH) with an influence  radius of 2--3 pc is mysterious. It presents a challenging puzzle for  our understanding of both star formation and dynamical interplays in the vicinity of a SMBH. In-situ formation of these stars requires cloud densities  which are about five orders of magnitude higher than the observed density in the GC \citep{Morr93};  a paradox which  amplifies at closer distances from the black hole. 

The young population in the GC appears to belong to different dynamical groups, with some fraction of the stars orbiting clockwise, and others counter-clockwise \citep{genz2000,levin03}. Between 20\% to 50\% of the young WR/O stars lie in a warped, clockwise rotating disk at radii $1'' < r < 12''$ \citep{Paum06,Bartk09,Lu09,Yelda14}. The young ages (4--6 Myr) of these mostly emission-line stars and their observed low eccentricities  \citep[e = 0.2--0.3, ][]{Bartk09, Lu13} probably favor an in-situ formation in a dense gas accretion disk that can overcome the SMBH tidal forces  \citep[e.g.][]{levin07,Alex08,bonn08}.

 The young stellar disk in the GC is observed to be truncated inside the central arcsecond.  Within a distance of 0.04 pc from the SMBH exists another  dense group of  somewhat fainter ($m_{K}=14-17$) and probably lower mass stars (``the S-stars"). These stars are so close to the black hole that tracing their individual orbits around  Sgr~A*, the radio source associated with the central SMBH,  is the key to determining the central mass and the distance to it \citep{Gill09,Boehle16,Gill17}. The S-stars are spatially isotropically distributed, and the orientation of  their orbits is consistent with a thermal distribution \citep{schoedel03,Gill09,Gill17}. First measurement of spectral lines of  S2, the brightest S-star, showed that this star appears to be  an early B-star \citep{ghe03,mar08} as it showed Brackett-$\gamma$ (Br$\gamma$) absorption and no CO absorption line. \cite{mar08} co-added 23.5 hours of observation of S2 and found that this star is an early B dwarf (B0--B2.5 V), eliminating the possibility for it to be the  core of a stripped giant star. Other S-stars may be lower mass B-stars, among which there
 are also late type stars. By co-adding the spectra (effective integration times of 20-80 minutes) of  five fainter S-stars, \cite{eis05} concluded that they are of the
 spectral type B4--B9. This allows for a wide range of main-sequence age between 6 and 400 Myr  for the S-stars.  
 
It is highly unlikely that the S-stars formed at their present location, since the SMBH's tidal forces are too strong to allow star formation at these distances. On the one hand, the sharp kinematic difference  between the O/WR stars  and the S-stars may suggest a different origin for these populations. On the other hand, there are two reasons one might consider these groups to be different components of the same parent population. First, since S-stars appear to be young, a formation near their current location  relaxes the constraints on the migration models. Second, the B-stars population extends (apparently) continuously beyond the central arcsecond and up to at least 0.5 pc from the SMBH \citep{Bartk10}. Several scenarios have been proposed to explain the origin of S-stars,  all of which can be assigned to one of three broad categories:  formation outside the central few parsecs followed by binary disruption \cite[e.g.][]{Hills,Perets07}, disk formation in the central parsec followed by a migration process \citep{levin07,Madigan09}, and rejuvenation of older stars by mergers or severe atmosphere alteration by envelope stripping \citep{Lee, genz03,davis05}.

A successful scenario should explain the young ages and randomly inclined orbits of the S-stars, via an efficient transport/relaxation
 mechanism. External formation followed by binary disruption is among the favorite scenarios as it also explains  some of the hyper-velocity stars that escape the Galaxy \citep[e.g.][]{brown09}. However, field binary disruption scenario  requires some minimum time span to explain thermal distribution of eccentricities in the S-star cusp \citep{Perets09}.  This may be incompatible with the shorter main-sequence lifetime of some S-stars. While it is plausible to assume that S-stars were initially formed in an outer stellar disk in the central parsec,  it has proved difficult  to find an explanation of how their original angular momenta  are strongly perturbed and later redistributed. Finally, many of the scenarios  that suggested environmental mechanisms to significantly alter the photosphere of in fact old S-stars \citep{Lee, Alex03,Hansen03,davis05}
either  have serious problems in explaining the S-stars' stellar properties or are not elaborated in detail. For example, it is not clear  how quickly after a major atmospheric perturbation a star will relax, and  whether such products experience significant mixing \citep[e.g.][]{Gaburov08}.

With all the aforementioned observational constraints and theoretical complexities the question of the origin and distribution of young stars in the GC has become one of the most remarkable issues in this field.  Here we perform  quantitative spectroscopy of S-stars and obtain their individual stellar  parameters. This enable us to locate these stars for the first time  in a Hertzsprung-Russell(HR)-like  diagram and provide new insights on their evolutionary stage and their puzzling origin. However, such an analysis demands a high signal-to-noise ($S/N>200$), comparable to that achieved in spectra of nearby Galactic OB-stars. In this study, we co-add more than 100 hours of adaptive-optics (AO) assisted spectroscopic observations of the GC over twelve years, to construct ultra deep spectra of eight S-stars. 

The paper is organized as follows: in section \ref{sec:obsandred} we list the observations used in this analysis and describe the reduction of the data to obtain deep spectra on a number of stars in the GC. Section \ref{sec:stellaratm} describes the stellar atmosphere models used to match the data and our fitting strategy . In  the result section (section \ref{sec:results}), we present the spectral and luminosity (gravity) classification of the S-stars as well as  parameters of individual stars based on the results of atmosphere model fits. In the final paragraph of the results section we compare the inferred stellar parameters with   stellar evolutionary models and discuss the inferred mass and age range for these stars. Section \ref{sec:discussion} discusses the uncertainties and considers our findings in the context of the GC environment. We present a short summary in the last paragraph of the same section.

%%%%%%%%%%%%%%%%%%%%%%%%%%%
\section{Observation and data reduction}
\label{sec:obsandred}

%%%%%%%%%%%%%%%%%%%%%%%%%%%
\subsection{Observations}
\label{sec:obs}

\begin{figure}
\begin{center}
\epsscale{1.1}
 \plotone{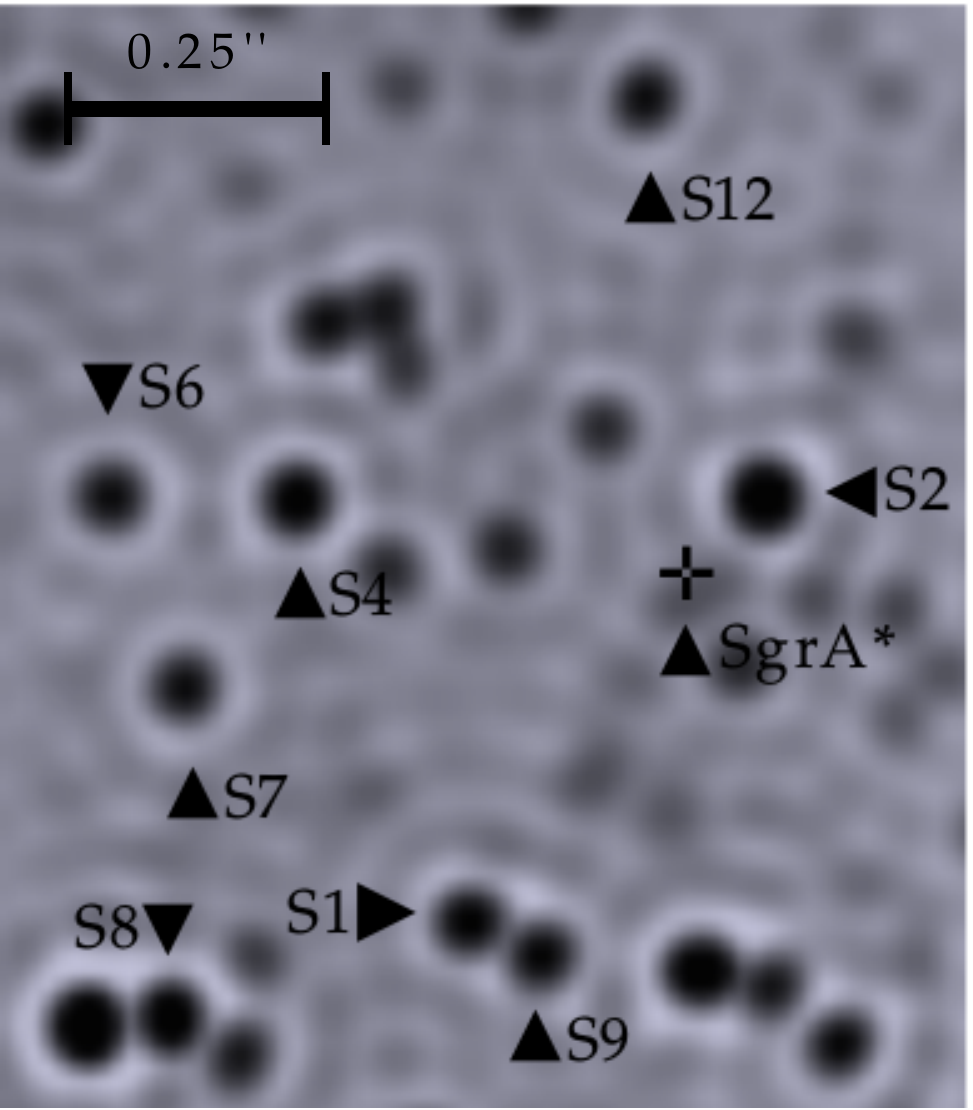}
\caption{The NACO  deconvolved image (K-band) of the central arcsecondimage, constructed from the April 5--18, 2017 data. The S-stars shown in the image are studied in this paper. S1, S2, S4, S6, S7, S8, S9, and S12 are the brightest ($m_{K}<15.5$) and possibly younger (no CO band detected) stars in the field. }
\label{fig:field}
\end{center}
\end{figure}

In this study we use twelve years (2004-2016) of  AO-assisted spectroscopic
data obtained with the integral field spectrometer SINFONI \citep{eis03a, bonn04} at UT4 of the VLT.  
A SINFONI exposure is reconstructed into a 3D data cube covering two spatial and one spectral dimension.
In this work, most of the data sets are observed using the grating covering the H+K bands with spectral resolution of $\sim$1500, and the exposure time of most individual frames is 600s. 
A few of the data sets observed between 2004 and 2006 use the K-band grating with a
spectral resolution of $\sim$4000 and have shorter exposures of 240-500 seconds per frame. 
In all data presented here, the spatial pixel scale used is 25 $\frac{mas}{pixel}$, which corresponds to a field of view (FOV) of $0.8'' \times 0.8''$.
Some of the observations presented in this work  have been published earlier in other studies e.g. \cite{eis03b,eis05}, \cite{Gill09}, and \cite{Gill17}. A summary of our data is given in
Table \ref{tab:summ_sinf}.

\begin{table}
\caption{Summary of SINFONI observations of the central arcsecond used for this study.\\
The on-target exposure times refer to S2. The actual exposure time for other stars may be lower 
due to the  dither pattern (centered around Sgr~A*). 
The FWHM is measured from a combined image of the star S2 on each date. The exposure time and the FWHM for one epoch (September 2-12, 2014) in which the S2 star is absent is determined from the star S4. 
\label{tab:summ_sinf}}
{\scriptsize 
\begin{center}
\begin{tabular}{l|ccc}
Date& Band &$t_\mathrm{exp}$ on S2 & FWHM \\
&&[min]&[mas]\\
\hline
July 15 2004 &H+K&10$\times$ 22  & 61  \\
July 17 2004 &K& 10$\times$ 3 & 78  \\
August 18--19 2004 &K& 10$\times$ 8 & 59  \\
March 19--20 2005 &K& 5 $\times$ 8  & 59 \\
June 18 2005 &K& 5$\times$ 16 & 60 \\
October 3--4 2005&H+K&8.3$\times$ 4& 72\\
October 5--7  2005&H+K&10$\times$ 9& 75\\
March 17 2006 &H+K& 8.3$\times$ 10 & 70  \\
August 16--18 2006 &H+K& 10$\times$ 8 & 70 \\ 
July 18--23 2007 &H+K& 10$\times$ 14& 70  \\
September 3--4 2007 &H+K& 10$\times$ 7& 68 \\
April 6 2008  &H+K& 10$\times$ 21 &56\\
May 21--25 2009 & H+K & 10$\times$ 11& 64\\
May 10--12 2010 & H+K & 10$\times$ 26& 58 \\
June 10--11 2010 & H+K & 10$\times$ 9& 63\\
July 08--09 2010 & H+K & 10$\times$ 3& 66\\
April 11--27 2011 & H+K & 10$\times$ 30& 55\\
May 2--14 2011 & H+K & 10$\times$ 11& 53 \\
July 27--August 3 2011& H+K & 10$\times$ 3& 64\\
March 18--20 2012 & H+K & 10$\times$ 4& 55\\
May 5--20 2012 & H+K & 10$\times$ 8& 55\\
Jun 26--01 July 2012 & H+K & 10$\times$ 28& 58 \\
July 07--11 2012 & H+K & 10$\times$ 32 & 64\\
April 5--18 2013 & H+K & 10$\times$ 121& 60\\
August 28--31 2013& H+K & 10$\times$ 26& 57\\
September 18--26 2013& H+K & 10$\times$ 19& 74\\
March 5--8 2014 & H+K & 10$\times$ 18& 59\\
March 28--April 10 2014 & H+K & 10$\times$ 33& 56\\
April 22--24 2014 & H+K & 10$\times$ 27& 54\\
May 8--9 2014 & H+K & 10$\times$ 29 & 54 \\
May 27--June 10 2014 & H+K & 10$\times$ 6& 102\\
July 08--20 2014 & H+K & 10$\times$ 33& 67\\
August 18--31 2014 & H+K & 10$\times$ 32& 70\\
September 2--12 2014 & H+K & 10$\times$ 37&  83\\
April 20--29 2015 & H+K & 10$\times$ 44& 63\\
May 18--23 2015 & H+K & 10$\times$ 9& 71\\
July 8--11 2015 & H+K & 10$\times$ 11& 76\\
April 14--16 2016 & H+K & 10$\times$ 19& 56 \\
July 9--11 2016 & H+K & 10$\times$ 3& 64 \\
\end{tabular}
\end{center}
}
\end{table}

%%%%%%%%%%%%%%%%%%%%%%%%%%%
\subsection{Data Reduction}
\label{sec:datared}

The standard data reduction for SINFONI data is applied to all the observations using 
the SINFONI data reduction package SPRED \citep{Schreiber04,Abuter06}.
On each image, we perform bad pixel correction, flat-fielding, and distortion correction. 
The wavelength calibration is performed using emission line gas lamps and then fine-tuned on the night-sky OH lines. 
For each frame, we subtract a sky image from the object image to eliminate emission features in the atmosphere. 
We then reconstruct the three-dimensional data cubes to produce an image of the instrument's FOV with a spectrum for each pixel.
Finally, the three-dimensional data cubes are corrected for atmospheric absorption using the telluric spectrum  of  
early-type stars  (Usually of spectral type B5) obtained in each respective night.
 
An individual 600s data cube gives a noisy spectrum, so the observed data are co-added in groups covering a \mbox{two-week} period. 
This short time scale boosts the S/N of the object, while at the same time ensuring that the Doppler-shifted spectra of fast moving stars have not changed significantly over the observing period. 
To increase the S/N for some of the fainter and slower stars, we allow combination of observed data sets within a two month period. 

The sample of S-stars studied in this work are selected to be among the brighter ($m_{K}<15.5$) and possibly younger stars (no CO bands) in the central arcsecond (see Fig. \ref{fig:field}).
We chose stars that are unconfused with other bright stars in at least few epochs between 2004-2016. 
From each combined cube, we extract a spectrum by manually selecting the source and background pixels. 
The background pixels are carefully selected to contain regions outside the mini-spiral emission in order to avoid nebular contamination in  Br$\gamma$. 
For fast moving stars like S2, the background selection is not a critical issue, as the the Br$\gamma$ absorption line of the star is Doppler-shifted by a large amount with respect to the nebular emission in the rest frame. For slow-moving sources the background selection is a delicate task. 
To check for source confusion in each epoch, we also use K-band photometric images observed with Naos-Conica \citep[NACO,][]{len98, rou98}. 
We only use spectra extracted from combined cubes in which 1) the star was  apparently unconfused with another star of a comparable or lower magnitude, and 2) we are able to visually identify the Br$\gamma$ absorption line.  

The source spectra are obtained by subtracting the weighted average of the background pixels from the weighted average of the source pixels. For spectra taken using the H+K grating, we oversample the spectra by a factor of two using a cubic spline interpolation method to reach 0.000245 $ \mathrm{\mu m}$ spectral pixel. The oversampling eases combining the  H+K data sets with spectral resolution of $\sim$1500 with the higher resolution (R$\sim$4000) spectra observed in the K-band grating during 2004-2006. It also allows fitting more accurate Doppler-shifts and consequently deriving a finer radial velocity grid.

%%%%%%%%%%%%%%%%%%%%%%%%%%%
\subsection{Baseline-finding and Normalization}
\label{sec:norm}

To normalize the spectra, the minimum component filter technique \citep{Wall97} is used.
The method is optimized for fitting a continuum amplitude to spectra with prominent emission/absorption lines. 
First, we mask the regions of the spectra where the absorption lines are evident. 
These patches include regions around Br$\gamma$ (2.166 $ \mathrm{\mu m}$), and \mbox{He I} (2.112 $ \mathrm{\mu m}$) in K-band, as well as Br13 (1.6109   $\mathrm{\mu m}$),  Br12 (1.6407   $\mathrm{\mu m}$), Br11 (1.6806   $\mathrm{\mu m}$), \mbox{He I} (1.7002   $\mathrm{\mu m}$), and  Br10  (1.7362   $\mathrm{\mu m}$) in H-band. 
A low order polynomial continuum is fitted to the masked spectra, and  is then subtracted as a first approximation of the continuum.
Next we perform a discrete Fourier transform of the resulting spectra, taper off the high frequencies using a Wiener low pass filter, and then apply an inverse Fourier transform. 
The result is a spectrum that contains only low-frequency modes. 
We obtain the final continuum array by adding the first approximation of it back to the result of the previous step. All spectra are then normalized using the obtained array.

%%%%%%%%%%%%%%%%%%%%%%%%%%%
\subsection{Radial Velocity Measurement and Combining the Spectra}
\label{sec:radialvel}

A Gaussian fit is used to locate the Br$\gamma$ line, which is the most prominent feature within the H- and K-band range.
Using the measured velocity from the location of the  Br$\gamma$ line, all the individual spectra are reverse Doppler-shifted to zero velocity. 
A preliminary combined spectrum is made with a S/N weighted average of all the individual spectra.  
The full preliminary combined spectrum is then used as a template which is cross-correlated with each individual spectrum to refine the estimate of the radial velocity  in every epoch. 
During this step, the individual spectra are over-sampled by a factor of four to achieve a more precise cross-correlation.

The cross correlation proceeds as follows: 
We create a grid of radial velocities with a spacing of 0.5 km/s, where the wavelength axis of the template is shifted using a proper Doppler-shift.
The shifted template is then linearly interpolated at the wavelength points of the individual spectrum in K-band to calculate the cross-correlation function (CCF). 
To find the radial velocity at which the global maximum of the CCF is located, a parabola is fitted to the points nearest to the maximum of the discrete CCF and the analytical maximum of the parabola is used.

\begin{figure}
\begin{center}
\epsscale{1}
\plotone{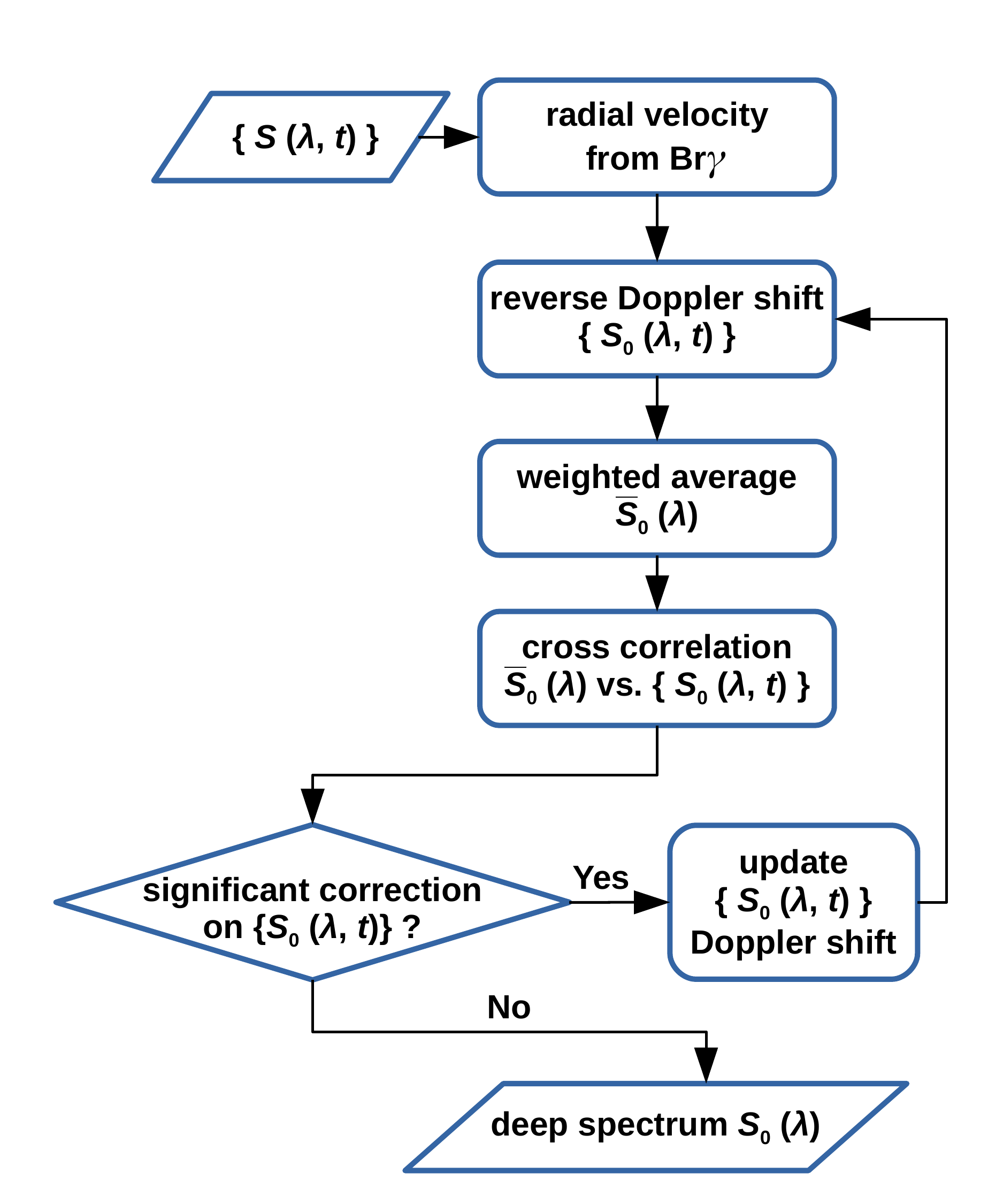}
\caption{Procedure of combining  spectra. The $\{S(\lambda,t)\}$ represents the observed spectra of the star at different times, while $\{S_0(\lambda,t)\}$ symbolizes
the zero-velocity versions of the spectra.}
\label{fig:flowchart}
\end{center}
\end{figure}

We correct the preliminary measurement of the radial velocities from each spectrum based on the results of the CCF. 
With the newly measured radial velocities, we create a new combined template spectrum. 
The CCF procedure is repeated iteratively to obtain ever more precise measurements of the radial velocity in each spectrum, and likewise an ever higher S/N combined spectrum. 
In case of the bright stars S2 and S4 this procedure is iterated until none of the individual radial velocity change by more than 1 km/s. 
For the fainter sources iterations continue until none of the individual radial velocities change by more than 7 km/s. 
At this point, the resulting template spectrum is considered the final deep combined spectrum.
Figure \ref{fig:flowchart} shows a schematic flowchart depicting all of the steps we use to construct the deep combined spectra. 

\section{Modeling} \label{sec:stellaratm}
In the light of previous similar studies on Galactic OB-stars and particularly GC objects \citep[e.g.][]{Najarro94,mar07}, we aim to determine  stellar parameters of the S-stars by comparing our observed high S/N spectra to our computed grid of atmosphere models. \cite{Repolust05} studied H and K band spectra of Galactic OB-stars and showed that the  stellar  parameters derived based on very high S/N (S/N $> 200$)  near-infrared spectroscopy  coincide in most cases with the optical ones within the typical errors\footnote{ The work of \cite{Repolust05} was based on the model atmosphere code FASTWIND \citep{Santolaya-Rey97}, but they also compared their calculations with the CMFGEN \citep{hm98} code and found a reasonable or  full agreement in most cases.}. 
\subsection{Stellar Atmosphere Models}

To determine the stellar parameters, we have computed atmosphere models using the code CMFGEN \citep{hm98}. This code solves the radiative transfer and rate equations iteratively to produce non-LTE atmosphere models. A spherical geometry is adopted to take into account the wind extension. Line-blanketing is included through the use of thousands of energy levels from H, He, C, N, O, Ne, Mg, Si, S, Ar, Ca, Fe and Ni. The solar composition of \citet{ga10} is adopted\footnote{\cite{mar08} used the \mbox{He I} lines at 2.149 and 2.184  $\mathrm{\mu m}$, and concluded that S2 is a He-rich star. In this study despite the 4.5 times longer integration time on the S2, these weaker lines are absent in the S2 spectrum. Therefore, a higher He abundance is not explored in this paper.}. A microturbulent velocity of 10 km $s^{-1}$ is used. The velocity structure is determined by the connection between the solution of the hydrodynamical equation in the inner atmosphere and a velocity law termed as $\beta$-velocity law in the external region ($v = v_{\infty} (1-R/r)^{\beta}$ with $v_{\infty}$ the maximum velocity at the top of the atmosphere and $R$ the stellar radius). The inner structure uses the radiative acceleration as computed from the level populations of the atmosphere model. It is iterated to provide a consistent solution. Once the atmosphere model converged, a formal solution of the radiative transfer equation is performed to yield the synthetic spectrum in the H and K bands. 

We have run a partial grid of models with effective temperatures between 19000 and 30000 K and $\log g$ between 3 and 4.5 (see the gray oval in Fig.\ref{fig:cmd_Tg}). This range brackets the values expected for main sequence early-B-stars (see Sec. \ref{sec:qualresults}). The projected rotational velocities with which the model spectra are convolved range from 0 to 400 km/s.

\subsection{Fitting Strategy}\label{sec:fitting}

At first, we  only include the \mbox{He I} lines  for a crude  measurement of the rotational velocity. Unlike hydrogen lines which are 
dominated by pressure broadening, rotational broadening dominates the \mbox{He I} line width at average to high projected rotational velocities\footnote{Sharp metal lines are most
sensitive to rotational broadening, however, they are few and very weak in B-stars, and are typically washed out due to higher rotational velocities.
We do not detect any metal lines in the spectra of young S-stars that are observed in this study. }. In particular, the \mbox{He I} 2.112/2.113  $\mathrm{\mu m}$ line is very helpful for estimating the rotational velocity since this line appear as a doublet comprising the He I triplet at 2.1120 $\mathrm{\mu m}$ ({$3p$ $^{3}{\rm P}^{\rm o}$} - {$4s$ $^{3}{\rm S}$}) and the He I singlet at 2.1132 $\mathrm{\mu m}$ ({$3p$ $^{1}{\rm P}^{\rm o}$} - {$4s$ $^{1}{\rm S}$}). With a line ratio of 0.5--0.6, the two lines appear as a shoulder feature if the projected rotational velocity is low, and if the rotational velocity is high enough we will detect a single line with a symmetric line profile. Such subtle changes in the \mbox{He I} line profile are traceable in our high resolution and high S/N spectra.

Next, we fit the spectrum in the determined sub-volume of parameter space to obtain temperature, surface gravity, and a modified estimate for rotational velocity $V \sin (i)$.
For this final fit we use  the \mbox{He I} lines in H- and K-band, Br10 and Br11 if they are not too
noisy, and Br$\gamma$ with the least weight. Our fitting relies less on the Br$\gamma$, as for this line there is a general mismatch between synthetic and observed profiles \citep[see Sec. 4.1 of ][]{Lenorzer2004}.
Contrary to Br$\gamma$, \mbox{He I} lines and Br10/11 are well understood and reproduced by synthetic spectra. 
The lines Br10/11 are excellent indicators of gravity, as line cores are deeper for lower gravities \citep{Repolust05}.

Before comparing the observed high S/N ratio  spectra with our synthetic spectral grid, the modeled spectra are convolved  with the  measured instrumental line-spread function\footnote{The average instrumental profile of SINFONI at Br$\gamma$ line using the H+K grating has a line spread which corresponds to a FWHM of $\sim 200$ km/s. The actual line profile of  SINFONI varies with wavelength and diffraction grating, and can have a complex shape with pronounced side-lobes \citep{iserlohe04, thatte12, george16, graeff16}. Working with  a very high S/N ratio spectrum in this study, it is necessary to  take into account the effect of the non-Gaussian instrumental profile. We use the average instrumental line profiles over the instrument's FOV at specific wavelengths close to the wavelength of the studied spectral lines.}. 
Afterwards, the model spectra, which now match the spectral resolution of our data, are convolved with a rotational broadening function for a range of rotational velocities.
Finally, each spectrum from the grid is compared to the observed spectrum of the star, and the chi-square is used to assess the goodness of fit.

The stellar parameters, i.e.  $T_{\mathrm{eff}}$, $\log g$, and $V \sin (i)$, of the model that gives the best fit to the observed spectrum, are the ``preliminary stellar parameters'' of the stars derived directly from the spectrum and the CMFGEN atmosphere models. These preliminary parameters are shown in Appendix \ref{prelim_param}. By comparing these three parameters, combined with photometric measurements, with the stellar evolution models, we derive the “final stellar parameters” for the S-stars (see Sec. \ref{stellar_param}).

%%%%%%%%%%%%%%%%%%%%%%%%%%%
\section{Results}\label{sec:results}
\subsection{Spectral and Luminosity (gravity) Classiffication} \label{sec:qualresults}

\begin{figure*}
\begin{center}
\plotone{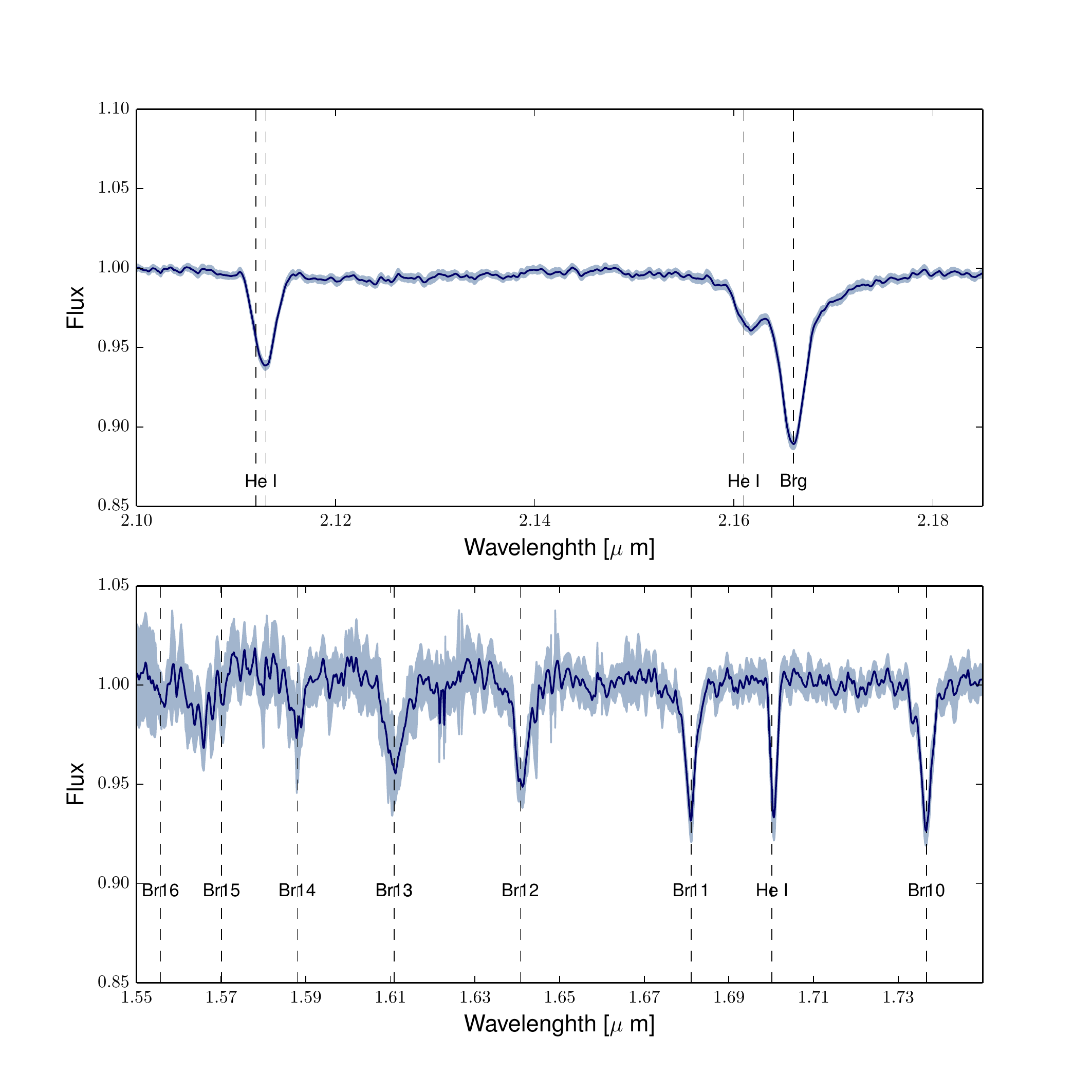}
\caption{Combined S2 spectrum from the 2004-2016 SINFONI data.
We see the Brackett series and three \mbox{He I} absorption lines, which are infrared characteristics of early B-type stars.  
The gray region illustrates the standard error of the mean calculated for combined epochs at a given wavelength. }
\label{fig:spec_s2}
\end{center}
\end{figure*}

\textbf{\textit{Tracing high gravity in the S2 spectrum}}: Using the procedure outlined in section \ref{sec:obsandred}, we combined $\sim 105$ hours of integration time in K-band and $\sim 99$ hours of integration time in H-band on the star S2. Figure \ref{fig:spec_s2} shows the combined spectrum in the K- and H-bands. The uncertainty of the spectrum is defined as the  standard error of the mean ($\frac{\sigma}{\sqrt N}$), derived from the standard deviation of the spectrum from the individual epochs. 

The resulting spectrum has a S/N of 480 in K-band. 
The most prominent K-band features are Br$\gamma$ and \mbox{He I} (at $2.112~\mu$m,  $2.161 \mu$m respectively) lines, in agreement with previous studies of S2 \citep{ghe03,eis05,mar08}. In our combined spectrum, we do not
see the weaker \mbox{He I} lines at 2.149 and 2.184  $\mathrm{\mu m}$ which have been claimed in \cite{mar08}. The \mbox{He II} line (2.189  $\mathrm{\mu m}$) is clearly absent in the S2 spectrum. Appendix~\ref{findfakelines} describes how we distinguish weak stellar lines from spurious features in the spectra. The \mbox{He I} line (2.058  $\mathrm{\mu m}$) lies  within a region of Earth's atmosphere where telluric $CO_{2}$ absorption is large. We can confirm that this line in S2 is not in strong emission as it is expected in stars with extended atmospheres \citep{Hansen96}.  Unfortunately,  due to telluric contamination this study is  inconclusive on \mbox{He I} (2.058  $\mathrm{\mu m}$) line profile.

Despite the higher extinction toward the GC at shorter wavelengths, we achieve a S/N of  $\sim~280$ in H-band. 
The main features of the  H-band spectrum of S2 are \mbox{He I} (1.7002  $\mathrm{\mu m}$) and higher order Brackett lines. 
We detect Br10--Br16 in the wavelength range of 1.53-1.75  $\mathrm{\mu m}$, of which four H I lines (Br13--Br16) are identified in an S-star for the first time.
The \mbox{He II} (12--7) line at 1.6918  $\mathrm{\mu m}$ is  not detected in the S2 spectrum. We also do not detect the \mbox{He II} (13-7) at 1.5719  $\mathrm{\mu m}$.
Due to the absence of the other \mbox{He II} lines, we assume that this line is absent as well, although it could be that we do not detect this line because it is close to near the Brackett (15--4) line at 1.5701  $\mathrm{\mu m}$.

By comparing the H- and K-band spectrum of S2 to near-infrared spectral atlases of O and B-stars \citep{hanson05, blum97}, we confirm the previous finding \citep{ghe03, mar08} that S2 is an early B-type star, i.e. 
B0-B3. Even with a very high S/N in both H and K-band we do not detect any \mbox{He II} lines, indicating that these lines are likely absent in the spectrum and therefore the temperature is below 30000 K.

Extending the infrared spectrum to H-band does not constrain the stellar subtype classification since the Brackett series  from Br10 to Br16 are present in all B-type stars \citep{blum97}. However, it provides an excellent gravity indicator (luminosity classification). 
As first found by \cite{Adams1922}, we observe stronger lines in the upper hydrogen series with increased
luminosity of the B-stars.  \cite{blum97} showed that while Br19 is confidently detected in the B supergiants,  Br15 or Br16 is the last of Brackett lines detected in B dwarfs. In high-gravity (dwarf) OB-stars, there is a deeper absorption in the wings of the higher 
order Brackett lines \citep[Fig.~12 of][]{hanson05}.

This behavior of hydrogen lines is well explained as resulting of the linear Stark profiles \citep{Struve1929} that are sensitive to the electron density. Quadratic Stark broadening is proportional to the fourth power of the upper principal quantum number. In hydrogen lines with relatively low upper principal quantum numbers, the core of the profile is dominated by Doppler broadeningو which is independent of electron density. \cite{Repolust05} show that for higher order Brackett lines, for example for Br10, the Stark width becomes considerable compared to Doppler-broadening. Consequently, even the line core becomes very sensitive to  the electron density. Therefore, the line cores of the upper hydrogen series become weaker with increasing gravity.

\begin{figure*}
\begin{center}
\epsscale{1}
\plotone{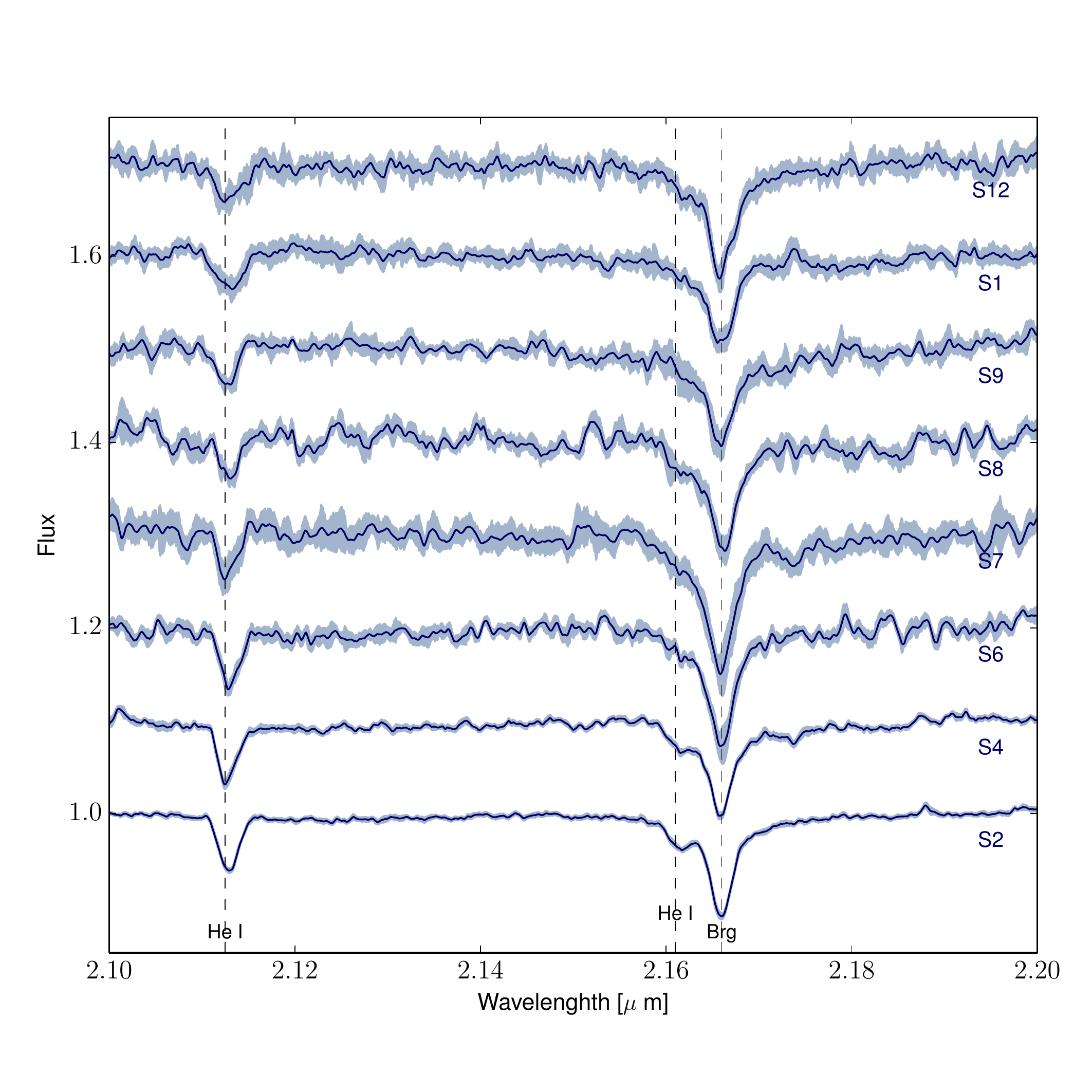}
\caption{Combined deep spectra of eight young S-stars ($K<15.5$ mag) in K-band. Individual Spectra observed between 2004-2016
are shifted to rest wavelength and combined. The gray region illustrates the standard deviation of combined epochs at a given wavelength.  
The names on the right side relate the spectra to the stars in Fig. \ref{fig:field}.
The most prominent absorption features are  Br$\gamma$ (2.166 $ \mathrm{\mu m}$) and two \mbox{He I} lines (2.112 $ \mathrm{\mu m}$ and 2.161 $ \mathrm{\mu m}$).
}
\label{fig:all_k}
\end{center}
\end{figure*}

In the S2 spectrum within the range of 1.53-1.75  $\mathrm{\mu m}$, we  detect  all the Brackett lines from Br10 to Br16. The H-band spectrum of S2  clearly shows a greater absorption in the wings of Br12--Br16 compared to Br10--Br11 (see lower plot of Fig.~\ref{fig:spec_s2}). We also observe  clearly weaker line cores in the higher order Brackett lines, which implies that the line profiles in S2
are resulting from a high electron density (high gravity). This is in agreement with a broad Br$\gamma$ in K-band which is merged with the \mbox{He I} (2.161  $\mathrm{\mu m}$) line; a morphology which is not expected for supergiants and giants (see Fig. 1 of  \cite{mar08} and upper panel of Fig.~\ref{fig:spec_s2} of this study).
Our H-band spectrum in shorter wavelengths ($\lambda < 1.57 \mathrm{\mu m}$) suffers from incomplete atmospheric correction. However, the observed decreasing strength of Br10 to Br16 indirectly provides another indication that the higher order  Brackett lines are either absent or very weak in the S2 spectrum, as is predicted for dwarf stars.
\emph{In summary, the new independent evidence from our high S/N H-band spectrum is that S2 is a high-gravity star.} 
\\
\textbf{\textit{Other S-stars appear as S2 siblings}}: By visual inspection, the spectra of all the observed young S-stars in this study are very similar (Fig. \ref{fig:all_k}).
In K-band the main stellar features are Br$\gamma$ (2.166  $\mathrm{\mu m}$), and \mbox{He I} (2.112  $\mathrm{\mu m}$).
K-band spectra of all of the observed S-stars exhibit considerable absorption at \mbox{He I} (2.112  $\mathrm{\mu m}$). This \mbox{He I} line appears in O7--O8 stars and remains visible  until B3 in dwarfs and as
late as B8--B9 in supergiants \citep{Hansen96}. Therefore, we conclude that all of the observed S-stars in this study  have a spectral type earlier than B3 if they are not supergiants.
Only for brighter stars, i.e. S2 and S4, we reach a S/N hight enough to detect all the Brackett lines between Br10-Br15 in H-band. For fainter stars we detect the \mbox{He I} line (1.7002  $\mathrm{\mu m}$), Br10, and Br11 in individual spectra. Similar to Br$\gamma$, none of the detected Brackett lines in H-band have narrow and deep morphology, which is expected in cooler supergiants. 

\begin{figure*}
\epsscale{1}
\hspace{-1.5cm}\includegraphics[trim={20 0 0 0},clip]{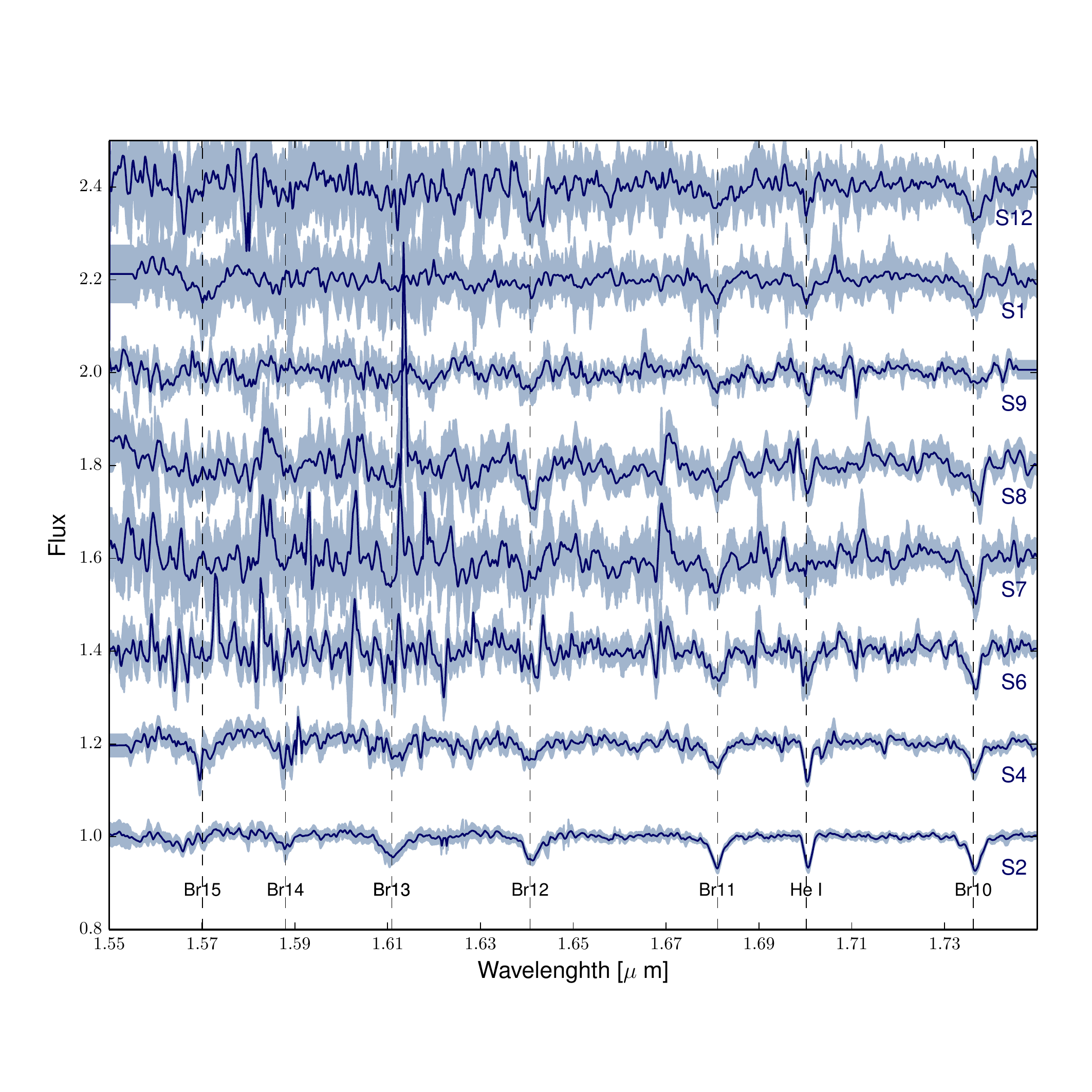}
\caption{Combined spectra of eight young S-stars ($K<15.5$ mag) in H-band. Individual Spectra observed between 2004-2016
are shifted to rest wavelength and combined. 
The gray region illustrates the standard error of the mean calculated for combined epochs at a given wavelength.  
The names on the right side relate the spectra to the stars in Fig. \ref{fig:field}.
The most prominent absorption features are \mbox{He I} (1.7002 $ \mathrm{\mu}$m) and  Brackett lines (Br10-Br15). }
\label{fig:all_h}

\end{figure*}

\cite{eis05} co-added the spectra of some fainter S-stars (K$\sim$ 15--15.8; S5, S08\footnote{This source was a confused source consisted of S51, S60, and S31 in July 2004.}, S12, S13, and S14) observed with an effective integration time of 20--80 minutes  using SINFONI. They measured an upper limit of 1${\buildrel _{\circ} \over {\mathrm{A}}}$ to the equivalent width (EW) of \mbox{He I} (2.112  $\mathrm{\mu m}$)  and $\mathrm{EW}>9$ \AA~ for Br$\gamma$ (2.166  $\mathrm{\mu m}$) line of these stars. An atlas of OB stars at 2$\mathrm{\mu m}$ (see Fig. 26, \citet{Hansen96}) provides $\mathrm{EW}\sim 0$  for the \mbox{He I} (2.112  $\mathrm{\mu m}$), and $\mathrm{EW} > 7$ \AA~ for Br$\gamma$ line of early B-stars. Taking into account the overall picture, \cite{eis05}  concluded that this group  has properties of \mbox{B4--9 V} stars. In our study, we include S12 from their sample to perform a quantitative spectroscopic analysis. As shown in Fig. \ref{fig:all_k} and \ref{fig:all_h} and is discussed in section \ref{stellar_param}, we find that S12 is an early B0--2.5V star. Although the noisy/confused spectra of S13, S5, and S14 do not allow for a full quantitative spectral analysis, their extracted  spectra from epochs in which they are relatively isolated reveal a distinct He I line in all of these stars. For the combined spectrum of S12 and S13, which are the brightest stars of the \cite{eis05} sample of faint stars, we measure an EW of 1.4 \AA~ and 5.6 \AA~ for \mbox{He I} (2.112  $\mathrm{\mu m}$) and  Br$\gamma$ lines respectively, which is consistent with these stars being early B dwarf-stars.

%%%%%%%%%%%%%%%%%%%%%%%%%%%
\subsection{Derived Stellar Parameters}\label{stellar_param}

For B-stars with negligible or thinner  winds, the main parameters which govern synthetic stellar atmospheric lines are $\log (g)$, $T_{\mathrm{eff}}$, and $V \sin (i)$. Lower effective temperatures, higher surface gravities, and higher rotational velocities can influence diagnostic lines in a similar way. Usually for OB-stars, $T_{\mathrm{eff}}$ is constrained from the strength of \mbox{He I}/\mbox{He II} or  Si III/Si IV lines. In the absence of \mbox{He II} and Si lines in the spectrum of the S-stars, it is hard to constrain their temperature. However, a very high S/N spectrum in K-band  and adding Brackett lines in H-band to our analysis comes to our rescue. \mbox{He I} (1.7002  $\mathrm{\mu m}$) and \mbox{He I} (2.112  $\mathrm{\mu m}$) lines are insensitive to $T_{\mathrm{eff}}$ for temperatures below 30000 K \citep[see Fig. 11,][]{Repolust05}. Moreover, the sensitivity of the \mbox{He I} lines on $\log (g)$ is only moderate at lower temperatures. Therefore, we can break the degeneracy of the rotational velocity, $V \sin (i)$, with the other two parameters. This is particularly possible since the \mbox{He I} lines at 2.112/2.113  $\mathrm{\mu m}$ appear as a doublet (see Sec. \ref{sec:fitting}). Therefore, we obtain the projected rotational velocities, $T_{\mathrm{eff}}$, and $\log(g)$ from the spectra. Moreover, we obtain the stellar luminosities  by combining the photometric and spectroscopic observations. First, we derive the absolute magnitudes by adopting a distance modulus of  $14.60$ based on the GC distance of $8.32$ kpc \citep{Gill17} and the K-band extinction of $A_{K}= 2.4$ \citep{Fritz11}.  The effective temperatures, which are determined from synthetic spectral fitting, are then applied to acquire the bolometric correction for the K-band magnitude, $BC_{K}$, adopting the relation 
\begin{equation} 
BC_{K}=-7.24 \times \log(T_{\mathrm{eff}}) +28.8
\end{equation}

from \cite{mar06}. Finally, the luminosity  of the star is calculated by 

\begin{equation} 
{\log{L}=-0.4\times(M_K-BC-M^{bol}_{sun}})
\end{equation}

with $M^{bol}_{sun}=4.74$ mag. The measured luminosities have relatively large uncertainties due to an unknown variable extinction towards the GC.

\begin{figure*}
\begin{center}
\epsscale{0.7}
 \plotone{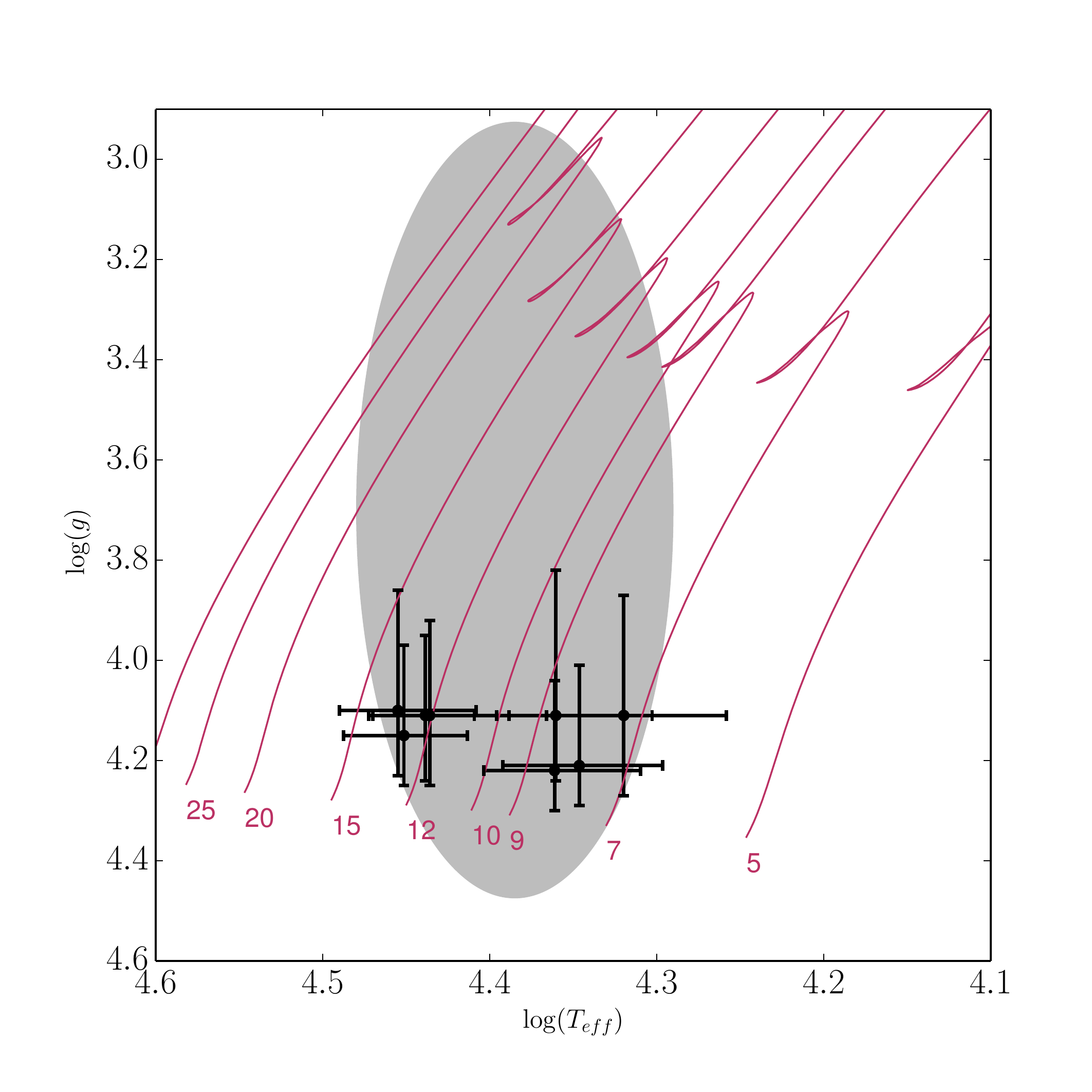}
\caption{The  fitted stellar parameters for the S-stars (black crosses) placed on the HR diagram in the  $T_{\mathrm{eff}}-log(g)$ plane. The Bonn stellar models for the Milky Way (MW) \citep{Brott11} of 5-30 $M_\odot$ stars are shown. The numbers below  the tracks denote the masses (in $M_\odot$) for which the evolutionary tracks are plotted.  The gray oval depicts parameters for which we have run models. Comparing models with the observed spectra shows that the S-stars are high-gravity OB-stars.}
\label{fig:cmd_Tg}
\end{center}
\end{figure*}

Once the two basic atmospheric parameters , for example, $T_{\mathrm{eff}}$ and $\log(g)$ are determined from the spectrum, one can estimate the age and mass of the stars by comparing the location of stars  in the $T_{\mathrm{eff}}$--$\log(g)$ plane with the stellar evolution tracks (see Fig. \ref{fig:cmd_Tg}). To exploit all the available observables and to achieve the best estimates for the fundamental stellar parameters, the comparison between observations and stellar models needs to be performed in a multidimensional space. Therefore, we would like to simultaneously match all the observables to stellar models while taking observed uncertainties  into account.  To that end, we use the Bayesian tool BONNSAI\footnote{The BONNSAI web-service is available at
\mbox{www.astro.uni-bonn.de/stars/bonnsai.}} \citep{Schneider14}. BONNSAI  is based on Bayes' theorem and determines probability distributions  to infer the fundamental stellar parameters from stellar models \citep{Brott11} and their uncertainties. The input observables to BONNSAI ($T_{\mathrm{eff}}$, $\log(g)$, $V \sin (i)$, and L) are derived from spectroscopic and photometric observations (preliminary parameters in Appendix \ref{prelim_param}). However, BONNSAI adjust these values within their observed uncertainties according to the parameters of the best-fitting model. 

The results of our analysis , i.e. the ``final stellar parameters''  with their respective uncertainties, are summarized in Table \ref{tabs:paramsTable}. For S2, we derive  $T_{\mathrm{eff}}= 28513^{+ 2388}_{- 2923}$ K, $\log(g) =  4.1^{+0.1}_{-0.2}$ (see appendix \ref{bonnsaiplots} for sample probability maps). These parameters put S2 in the hottest B-stars  group, most probably a B0 star. With $\log (g)$=4.1, S2 is  a dwarf star  \citep[$\log (g) > 3.7-4.2$;][]{trundle07,Nieva14}. This confirms  previous findings by \cite{mar08} and our new evidence from the H-band spectrum (see Sec. \ref{sec:qualresults}). The obtained fitted stellar parameters for other S-stars investigated here are very similar. Therefore, in the following we discuss their comprehensive properties. 

\textbf{\textit{Temperature}}: The temperatures obtained for the eight stars analyzed here lie within the range of \mbox{$21000-28500$ K}. This temperature range is characteristic of spectral types from B0 to B2.5.  The errors for the effective temperatures are on the order of $2000$ K, which corresponds to $10\%$ at the lowest fitted temperature. The uncertainty of $\sim$2000 K is derived after employing BONNSAI, which accounts for the additional information of correlation between temperature and other observed parameters through stellar evolution models. Therefore, we can re-estimate the temperature, within the observed uncertainties, with higher precision. The error bars on temperatures that are inferred directly from atmosphere modeling, i.e. by marginalization over  $\log (g)$ and $V \sin (i)$, are as high as  $\sim 3500$ K (see Appendix \ref{prelim_param}). This is not surprising, considering that there are no direct temperature tracers available in spectra of these stars.

\textbf{\textit{Surface gravity}}:  Comparing the derived $\log (g)$ values with two sets of different evolutionary tracks and also some observational studies we can confirm that $\log(g)$ for S1, S2, S4, S6, S7, S8, S9, and S12 are  within the dwarf range \citep[$\log (g) > 3.4-4.2$;][]{trundle07,Nieva14,geneva12,Brott11}. Figure \ref{fig:cmd_Tg} also shows that the observed S-stars lie within the main sequence stage, i.e below the turn due to central hydrogen exhaustion in the stars. The uncertainties for the surface gravities derived  are on the order of 0.2-0.3 dex. As the fitted $\log(g)$ decreases, the uncertainty for surface gravities increases.

\textbf{\textit{Rotational velocity}}: The projected rotational velocities derived fall in the range from 60 to \mbox{170 km/s}. The error for stars with  $V \sin (i) \sim 100$ km/s, i.e. S2, S4, S8, and S9,  is $\sim$50 km/s. For stars with \mbox{$V \sin (i) \sim 150-170$ km/s}, i.e. S1, S6, and S12, the uncertainty reaches  70 km/s. The star S7 shows a line profile consistent with a relatively  low $V\sin(i)$ of 60 km/s.  The uncertainties in $T_{\mathrm{eff}}$ and $\log(g)$ have little effect upon our estimated projected rotational velocities. 

Typical values of projected rotational velocities in B-type stars in the Galaxy are around  100 km/s, which extends to values higher than 200 km/s (e.g. \cite{Abt,Levato13}). Some studies suggest  that  the $V \sin (i)$ distribution is overall flat within 0-150 km/s \citep{Bragan12}, while others report observing a bimodal distribution, containing a slow group with a $V \sin (i)$ peak near 20 km/s \citep{Garmany15}. Disregarding details, the measured rotational velocities of S-stars are well within the observed distribution of  $V \sin (i)$ for B-type stars in the Milky Way.

\textbf{\textit{Stellar luminosity}}:  We derive a luminosity range of \mbox{$\log (L/L_\odot)\sim$ 3.6--4.3} for our sample of S-stars. The uncertainties for
the derived \mbox{$\log (L/L_\odot)\sim$} are on the order of 0.2-0.3 dex. These large uncertainties, are conservative estimates to take into account the photometric errors, uncertainty
of 0.2 on K-band extinction, and 0.2 kpc on the distance to the GC.

\begin{figure*}
\begin{center}
\epsscale{0.7}
 \plotone{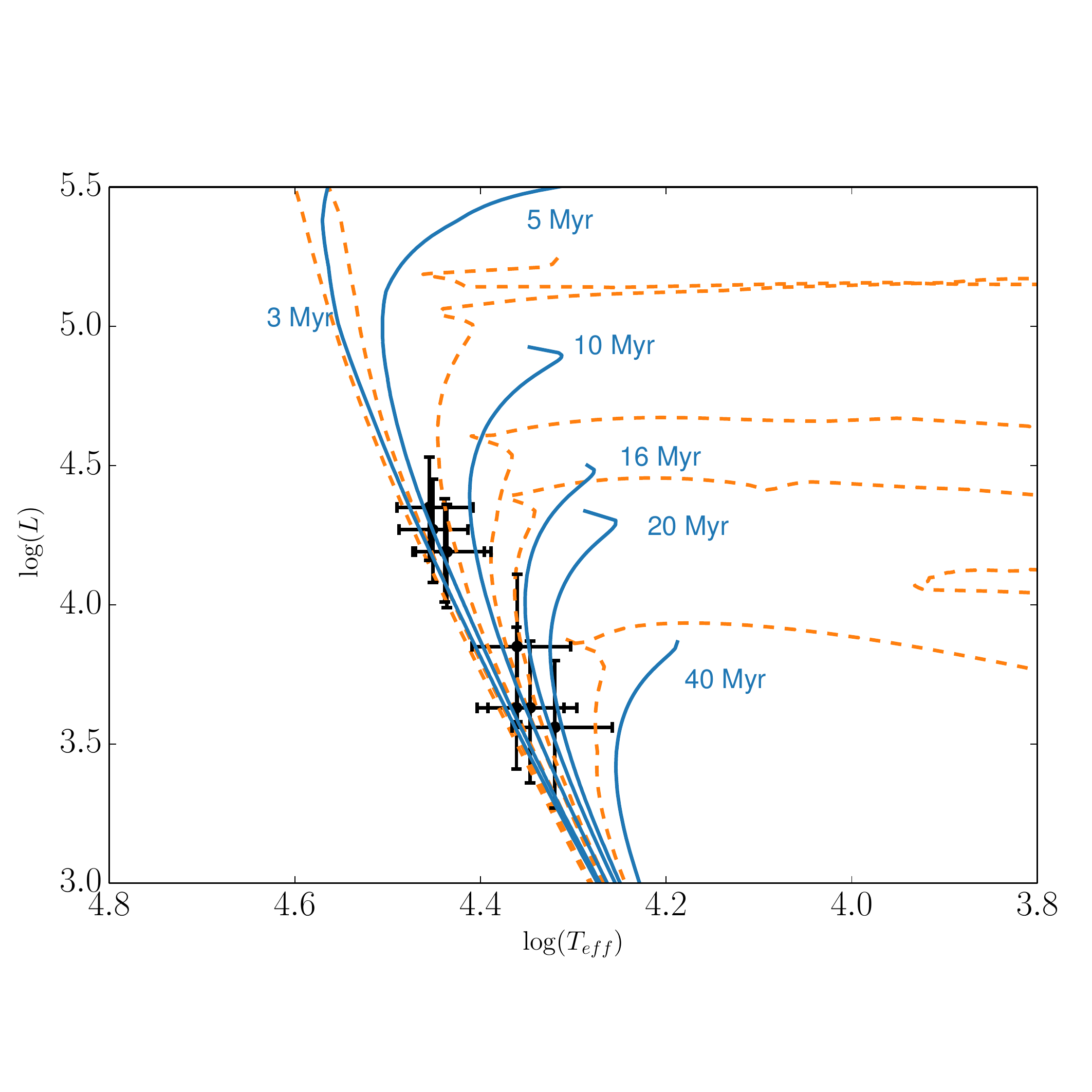}
\caption{The black crosses mark the  fitted stellar parameters for the S-stars on the HR diagram in the $T_{\mathrm{eff}}-log(L)$ plane. The Bonn stellar models for the Milky way \citep{Brott11} and  solar metallicity rotating Geneva isochrones \citep{geneva12} of 3-40 Myr are shown in respectively solid blue and dashed orange lines. The two sets of models  agree within the inferred uncertainties.}
\label{fig:cmd_LT}
\end{center}
\end{figure*}

\textbf{\textit{Mass and age:}} 
The location of the studied S-stars in the $T_{\mathrm{eff}}$--$\log(g)$ plane is shown in Fig. \ref{fig:cmd_Tg}, where also evolutionary tracks \citep{Brott11} are also illustrated for comparison. Figure \ref{fig:cmd_LT}  shows the theoretical isochrones together with the S-stars. Although such  diagrams are helpful to understand the evolutionary stage of the S-stars, one should not forget that the estimated masses and ages presented in Table \ref{tabs:paramsTable} are not derived  from  a simple comparison with theoretical isochrones in a 2-parameter plane.

Through a multidimensional comparison of observables and models with BONNSAI, we calculate the posterior probability distribution of the initial mass for each star\footnote{The inferred initial and actual masses from \cite{Brott11} models are not different.}, yielding its mean mass and associated 68\% confidence intervals (see Appendix \ref{bonnsaiplots}). The observables on which the mass and age estimates are based are luminosity, $\log(g)$, $T_{\mathrm{eff}}$, and $V \sin (i)$.

We find an evolutionary initial mass of $ 13.6^{+2.2}_{-1.8}$ $M_\odot$ for S2. The mass of the young S-stars in our study are within $\sim$8--14$M_\odot$. We derive an age of  $6.6^{+3.4 }_{-4.7 }$ Myr for S2. Correspondingly, we estimate the age of the other investigated S-stars to be less than 15 Myr, which is compatible with the age of S2 within the error bars. Moreover, we can firmly exclude ages that exceed 25 Myr for the S-stars studied in this work. This upper age limit is derived by considering the highest age found for these S-stars and accounting for the highest uncertainty associated with it.

\section{Discussion and Summary} \label{sec:discussion}
The masses presented in Table \ref{tabs:paramsTable} are spectroscopic-evolutionary masses, which are derived employing all available observations and stellar evolution models.
In addition to these masses, it is possible to estimate pure spectroscopic masses. If the distance and extinction of the star is  known, then photometry yields the star's absolute bolometric luminosity. This allows the radius of the star ‌to be  derived from  the Stefan-Boltzmann law. Using the radius of the star and the surface gravity of the best matching stellar atmosphere model, the star's mass can, in principle, be found employing $M = g R^{2}/G$.  The calculated spectroscopic masses of S-stars are lower than the evolutionary masses. For S2, we derive a spectroscopic mass of $9 M_\odot$, which in this case is $\sim 25\%$ lower than the spectroscopic-evolutionary mass of S2. \cite{mar08} analyzed the first deep spectrum of S2, and concluded that S2 is a genuine massive star. Although the qualitative conclusion of their study is valid, the calculated spectroscopic masses, presented in their table 1, should be corrected to lower values due to a calculation mistake (Martins, private communication). It is important to note the large errors in spectroscopic mass  estimates.  No error bars are calculated in \cite{mar08}, however, we do not expect lower uncertainties compared to this work since their study is based on 4.5 times lower observation time. Our inferred uncertainty of $\sim 0.3$ dex in the $\log(g)$ estimate  corresponds to $10^{0.3}\sim 2$ factor in the spectroscopic mass and hence such estimates are of limited value.  In other words, the calculated spectroscopic masses both in this study and \cite{mar08} are highly uncertain ($\sim200$\%) and are lower than spectroscopic-evolutionary masses.

\cite{Herrero} pointed out a mass discrepancy issue for massive stars: the stellar masses derived by spectroscopic methods are systematically smaller than those inferred from evolutionary models. ‌Besides the known mass discrepancy issue, large uncertainties in surface gravity estimation of S-stars have strongly affected the spectroscopic mass estimation. Although individual spectroscopic masses are highly uncertain, we perform an experiment in which we used the uncertain spectroscopic mass of S2 as one of the available observables with which we estimate the evolutionary stage of the star. We  find that the normal main sequence stellar models are unable to explain this particular set of observables with a confidence of 95\%.

As opposed to spectroscopic masses which are very sensitive to $\log(g)$, our spectroscopic-evolutionary masses are inferred using stellar evolution models and all the available observables including their uncertainties simultaneously.  In this way, we take into account the additional information of correlation between stellar parameters  through stellar evolution models so that stellar parameters can be deduced with higher precision. The error bars of the spectroscopic-evolutionary masses (see the last paragraph of Sect. \ref{stellar_param}) are lower by a factor of 10 than those for spectroscopic masses. We have derived a mass range of 8--14 $M_\odot$  for the S-stars in this study (see sect.  \ref{stellar_param}). Although one can interpret the discrepancy between the spectroscopic and spectroscopic-evolutionary masses as a hint of non-standard evolution of the S-stars, we think the high uncertainty on spectroscopic masses hinders such conclusions. 

For our best studied S-star, S2, we derive an age of $6.6^{+3.4 }_{-4.7 }$ Myr. This value is compatible with the estimated age of the clockwise rotating disk of young stars in the GC. We calculate the age of the other studied S-stars to be less than 15 Myr, and we can firmly reject ages which exceed 25 Myr for these S-stars. Within the uncertainties, their inferred ages are consistent with the age of S2. In other words, the inferred ages for the investigated S-stars favor the scenario in which these stars are  members of a recent star formation episode that possibly also created the observed young stellar disk or similar dissolved disks in the GC.

The proposed scenarios to explain the origin of S-stars can be classified to one of three wide categories:  formation outside the central few parsecs (field binaries) followed by binary disruption \cite[e.g.][]{Hills,Perets07}, formation in the central parsec, in the O/WR-disk(s), followed by a migration process, and rejuvenation of older stars by mergers, tidal heating, or envelope stripping \citep{Lee, genz03,davis05}. If the S-stars initially were formed in an outer stellar disk in the central parsec, their original angular momenta must have been strongly perturbed and later redistributed. \cite{levin07}  suggested that the S-stars are initially formed on the  stellar disk and migrated  through  type-I migration in planetary dynamics within a timescale of $10^5$ yr. Then they invoke \textit{resonant relaxation} \citep[RR,][]{Rauch96} to redistribute angular momenta into an isotropic distribution. Even their shortest RR timescale, i.e. 40 Myr, is  longer than our inferred ages for S-stars. However, these time scales might reduce to a few Myr in the presence of  an intermediate-mass black hole in  a close elliptical orbit \citep{merritt09}. The simulations of \cite{Antonini13} discussed  RR times in the context of the post-capture dynamical evolution of binaries. They  also achieved  shorter RR time if there is a dense cluster of 10$M_\odot$ black holes. Nonetheless, it is important to note that to date no intermediate-mass black holes have been found in the GC. This issue is common with field-binary scenarios,  as intermediate massive black holes are among the best candidates to conduct as massive perturbers and provide faster relaxation for older-field-binary disruption scenarios.

Numerical simulations by \cite{Madigan09} show that it is possible that young binary stars in the disk(s) migrate into the vicinity of the  SMBH and are then  tidally broken up via the \cite{Hills} mechanism. \cite{Madigan14} explored formation scenarios for the B-stars within 1 pc in the GC and found that the preferred scenarios correlate strongly with the magnitude interval. They found that stars with \mbox{$14<m_{K}<15$} (S-stars in their sample) match a dissolved disk origin formed in an episodic in-situ star formation scenario suggested  by \cite{Seth06}. In this scenario the disk is  puffed up  due to gravitational interactions between stars and  the  gravitational  influence  of  the  circumnuclear  disk.

 Although the proposed scenarios so far show that a disk origin for S-stars is possible, it is unclear whether the necessary conditions predicted by different scenarios are  fulfilled in the GC. Our future explanation of the S-stars should account for the following findings revealed by this long-term spectroscopic monitoring of the S-stars:

\textbf{ 1.} The deep spectra of S2 (K-band S/N: 480 and H-band S/N: 280), S4,  and also combined spectrum  of other fainter S-stars, disclose a 
clear Stark broadening in the Brackett lines which implies high surface gravity of S-stars. This finding is established by our detailed stellar atmospheric+evolutionary model analysis which employs line profiles of the complete Brackett series (excellent indicators of gravity) in H- and K-band. 

 \textbf{2.} We measure the rotational velocities by analyzing the \mbox{He I} lines whose line width is mainly due to stellar rotation. In particular we use the \mbox{He I} 2.112/2.113 $ \mathrm{\mu m}$ doublet which  exhibits a different line profile at low and high velocities, traceable in our high S/N data. The observed rotational velocities of studied S-stars are within 60-170 km/s.

\textbf{3.} High quality spectra and good tracers for rotational velocity and surface gravity allow us to estimate the temperature of stars only based on the line profile of Brackett and \mbox{He I} lines. Our stellar atmosphere+evolution model analysis yields a temperature range of 21000-28500 K. This higher inferred temperature range classifies the analyzed S-stars as type B0-B2.5. This finding is  confirmed with quantitative analyses of the spectra which show a considerable absorption at \mbox{He I} (2.112 $ \mathrm{\mu m}$) for all of the investigated S-stars.  

\textbf{4.} Our high S/N spectra for S2 and seven fainter S-stars confirm the previous findings on the apparent normalcy of S-star spectra \citep{ghe03, eis05,mar08}.

\textbf{5.} The inferred masses of the S-stars analyzed, 8--14$M_\odot$, is consistent with the bright S-stars being genuine main-sequence massive stars. We also infer the evolutionary stage of the S-stars by comparing our photometric and spectroscopic observations simultaneously  to stellar evolution models. We derive an age of  $6.6^{+3.4 }_{-4.7 }$ Myr for S2. With  higher uncertainties, we estimate the age range of the other studied S-stars to be less than 15 Myr. The inferred ages for all the analyzed S-stars are compatible with the age of the young clockwise rotating stellar  disk in the GC. Our analysis excludes ages which exceed 25 Myr for these S-stars.

\begin{table*}
\caption{Stellar parameters of the S-stars. The parameters were determined by fitting detailed model atmospheres to spectra and then by employing all available observables to compare with stellar evolution models. The errors quoted in this table are from the CMFGEN+BONNSAI analysis and are derived from the full posterior probability distribution of the model parameters. The last two columns give the spectral type and the K-band magnitude for each star. }
\label{tabs:paramsTable}
{\scriptsize
\begin{center}
\begin{tabular}{lcccccccccc}
Star&$T_{\mathrm{eff}}$[K]&$\log(g)$(cgs)&$\log (L/L_\odot)$& $R/R_\odot$& $Mass/M_\odot$& Age[Myr]& $V \sin (i)[km s^{-1}]$&SP&$m_K$ \\
\hline
\\
S1  &$ 27450^{+ 2239}_{- 2569}$&$ 4.11^{+0.13}_{-0.16}$&$ 4.19^{+ 0.19}_{-0.18}$& $5.19^{+1.13}_{-0.76}$&$ 12.40^{+2.0}_{-1.7}$&$  4.3^{+4.2 }_{-4.2 }$&$ 150.00^{+78}_{-44}$&B0--B3&14.8&\\
S2  &$ 28513^{+ 2388}_{- 2923}$&$ 4.10^{+0.13}_{-0.24}$&$ 4.35^{+ 0.18}_{-0.19}$& $5.53^{+1.77}_{-0.79}$&$ 13.60^{+2.2}_{-1.8}$&$  6.6^{+3.4 }_{-4.7 }$&$ 110.00^{+68}_{-45}$&B0--B3&14.1&\\
S4  &$ 27288^{+ 2224}_{- 2823}$&$ 4.11^{+0.14}_{-0.19}$&$ 4.19^{+ 0.17}_{-0.20}$& $5.23^{+1.21}_{-0.85}$&$ 12.20^{+1.9}_{-1.7}$&$  5.9^{+4.1 }_{-5.5 }$&$ 110.00^{+67}_{-45}$&B0--B3&14.6&\\
S6 &$  22932^{+ 2725}_{- 2851}$&$ 4.11^{+0.13}_{-0.29}$&$ 3.85^{+ 0.26}_{-0.27}$& $4.51^{+1.79}_{-0.78}$&$  9.20^{+1.9}_{-1.5}$&$ 11.5^{+7.8 }_{-9.3 }$&$ 170.00^{+84}_{-46}$&B0--B3&15.4&\\
S7  &$ 22972^{+ 2353}_{- 2565}$&$ 4.22^{+0.08}_{-0.18}$&$ 3.63^{+ 0.29}_{-0.22}$& $3.89^{+0.99}_{-0.53}$&$  8.40^{+1.7}_{-1.4}$&$  0.3^{+14.3}_{-0.3 }$&$  60.00^{+56}_{-41}$&B0--B3&15.3&\\
S8  &$ 28274^{+ 2461}_{- 2374}$&$ 4.15^{+0.10}_{-0.18}$&$ 4.27^{+ 0.18}_{-0.19}$& $5.11^{+1.30}_{-0.59}$&$ 13.20^{+2.2}_{-1.8}$&$  3.1^{+4.4 }_{-3.0 }$&$ 100.00^{+68}_{-54}$&B0--B3&14.5&\\
S9  &$ 22202^{+ 2463}_{- 2416}$&$ 4.21^{+0.08}_{-0.20}$&$ 3.63^{+ 0.24}_{-0.27}$& $3.89^{+1.04}_{-0.55}$&$  8.20^{+1.5}_{-1.4}$&$  3.5^{+13.2}_{-3.5 }$&$ 100.00^{+66}_{-51}$&B0--B3&15.1&\\
S12 &$ 20882^{+ 2348}_{- 2760}$&$ 4.11^{+0.16}_{-0.24}$&$ 3.56^{+ 0.24}_{-0.29}$&$4.07^{+1.31}_{-0.78}$&$   7.60^{+1.4}_{-1.3}$&$ 14.7^{+10.6}_{-13.9}$& $150.00^{+79}_{-47}$&B0--B3&15.5&\\
                                                                                                                                                                               
\end{tabular}
\end{center}
}
\end{table*}

%%%%%%%%%%%%%%%%%

\acknowledgments
Acknowledgments.
We thank John Hillier for making CMFGEN available to the community and for constant help with it.
We would like to thank Joachim Puls and Wolfgang Brandner for helpful discussions. We also thank our referee, whose report helped in improving the manuscript.
\clearpage
\appendix
\section{A. Investigation of weak stellar features}
\label{findfakelines}

\begin{figure}[h]
\begin{center}
\epsscale{1.1}
 \plotone{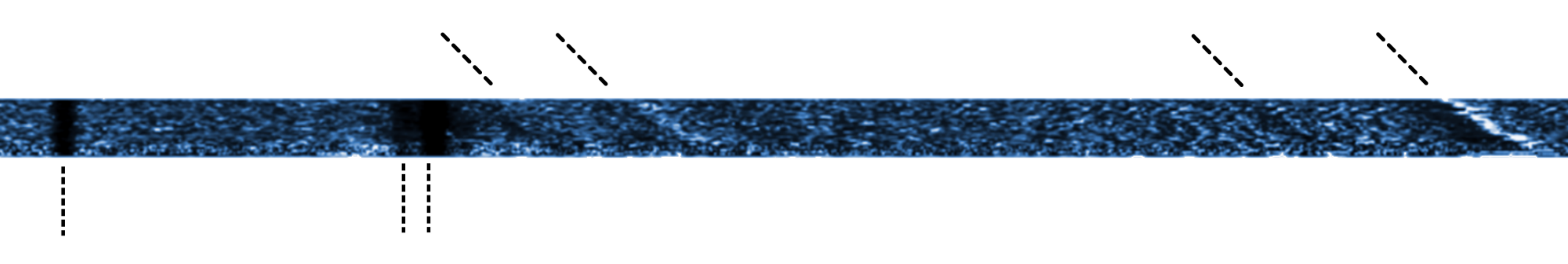}
\caption{A sample stacked spectra at the stellar rest-frame. Each horizontal stripe represents one epoch of stellar spectra in a selected region of K-band. The inclined features with non-zero velocity relative to the stellar rest-frame are  instrumental artefacts or residuals of  the incomplete atmospheric correction.}
\label{fig:findfake}
\end{center}
\end{figure}

In order to distinguish between weak stellar lines and residual artefacts in the spectra (either instrumental artefacts or residuals of  the incomplete atmospheric correction), we stack the spectra at rest-frame vertically on a map so that each horizontal stripe represents one epoch of stellar spectra. 
Stellar features appear as vertical lines on such a map, while artefacts will appear as inclined lines due to their non-zero velocity relative to the stellar rest-frame as the star changes velocity throughout its orbit. We use such maps as diagnostic plots to distinguish weak stellar lines from artefacts. In Fig. \ref{fig:findfake}, we show a sample map in K-band which reveals some non-stellar features as inclined lines in the stacked spectra.

\section{B. Preliminary stellar parameters}
\label{prelim_param}

The preliminary stellar parameters of the stars derived directly from the spectrum and the CMFGEN atmosphere models, are given in the table below. The stellar parameters, $T_{\mathrm{eff}}$, $\log g$, and $V \sin (i)$, are of the CMFGEN model that gives the best fit to the observed spectra, and the luminosities are derived by combining photometric and spectroscopic observations. Please note the large uncertainties on the inferred preliminary stellar parameters which is partly due to lack of direct tracers of temperature in these spectra. Since the observables are correlated, the large uncertainty in temperature determination affects the determination of other parameters. To derive the final stellar parameters (presented in the Table \ref{tabs:paramsTable}), we benefit from combining the spectroscopic and photometric observations and also accounting for the additional information of correlation between these parameters through stellar evolution models.

\begin{table*}[h]
\caption{Preliminary stellar parameters of the S-stars. The parameters were determined by fitting model atmospheres generated with the CMFGEN  to the observed spectra combined with photometric measurements.}
\label{tabs:paramsTable}
{\scriptsize
\begin{center}
\begin{tabular}{lcccccccccc}
Star&$T_{\mathrm{eff}}$[K]&$\log(g)$(cgs)&$\log (L/L_\odot)$& $V \sin (i)[km s^{-1}]$\\
\hline
\\
S1  &$ 30000\pm3500$&$ 3.8\pm0.45$&$ 4.17\pm0.2$&$ 140\pm50$\\
S2  &$ 30000\pm3600$&$ 3.8\pm0.40$&$ 4.37\pm0.2$&$ 100\pm50$\\
S4  &$ 30000\pm3700$&$ 3.8\pm0.45$&$ 4.18\pm0.2$&$ 100\pm50$\\
S6  &$ 27000\pm3600$&$ 3.4\pm0.56$&$ 3.80\pm0.3$&$ 160\pm50$\\
S7  &$ 26000\pm3700$&$ 4.3\pm0.46$&$ 3.7 \pm0.3$&$  55\pm70$\\
S8  &$ 30000\pm3700$&$ 4.1\pm0.45$&$ 4.3 \pm0.2$&$  85\pm50$\\
S9  &$ 24000\pm3700$&$ 4.2\pm0.35$&$ 3.7 \pm0.3$&$  90\pm50$\\
S12 &$ 23000\pm3600$&$ 3.6\pm0.55$&$ 3.6 \pm0.3$&$ 140\pm50$\\
                                                                                                                                                                      
\end{tabular}
\end{center}
}
\end{table*}

\section{C. Example posterior probability maps. }
\label{bonnsaiplots}

\begin{figure}[h]
\begin{center}
\includegraphics[trim=0 0 0 0,clip,width=89mm]{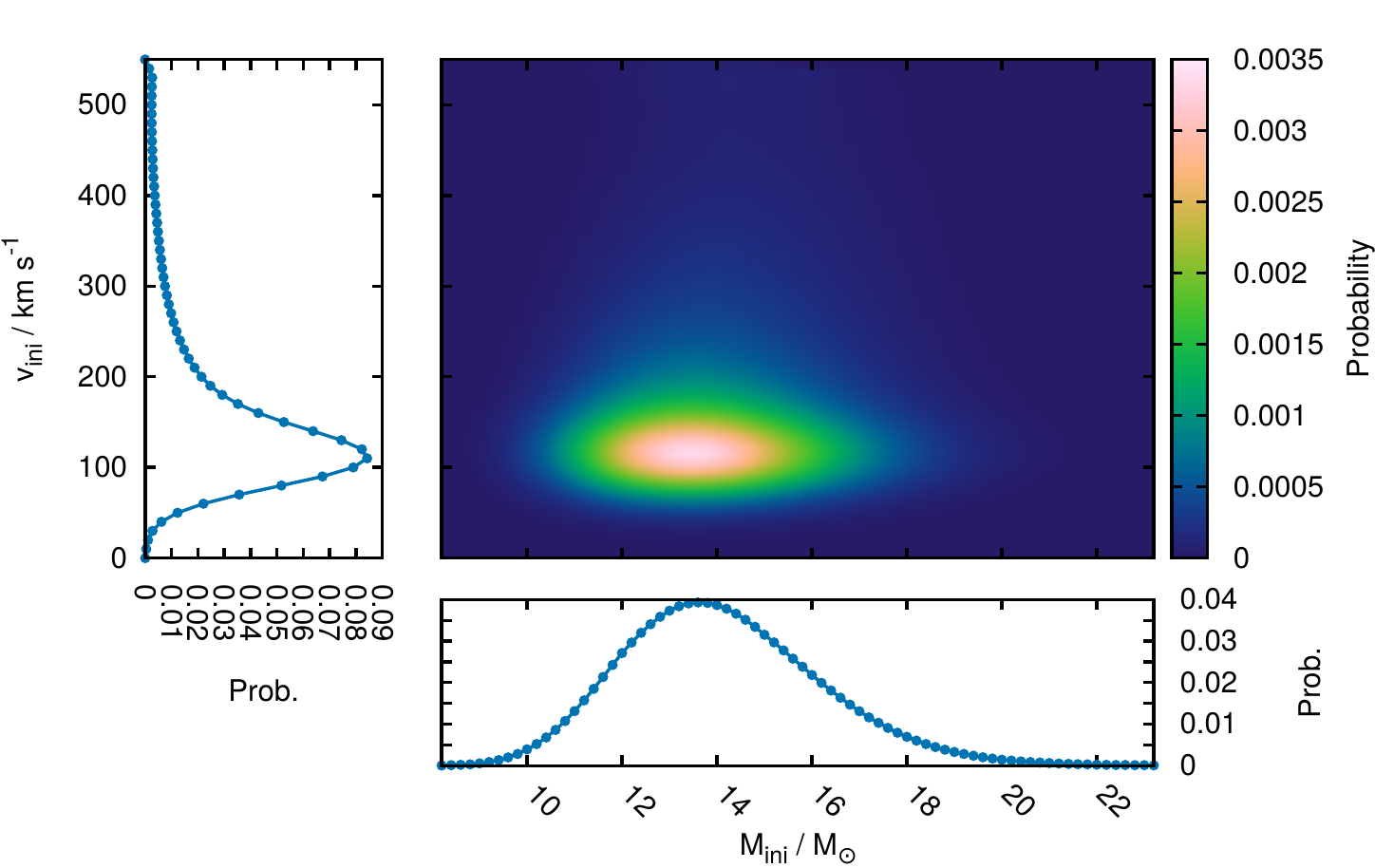}
\includegraphics[trim=0 0 0 0,clip,width=89mm]{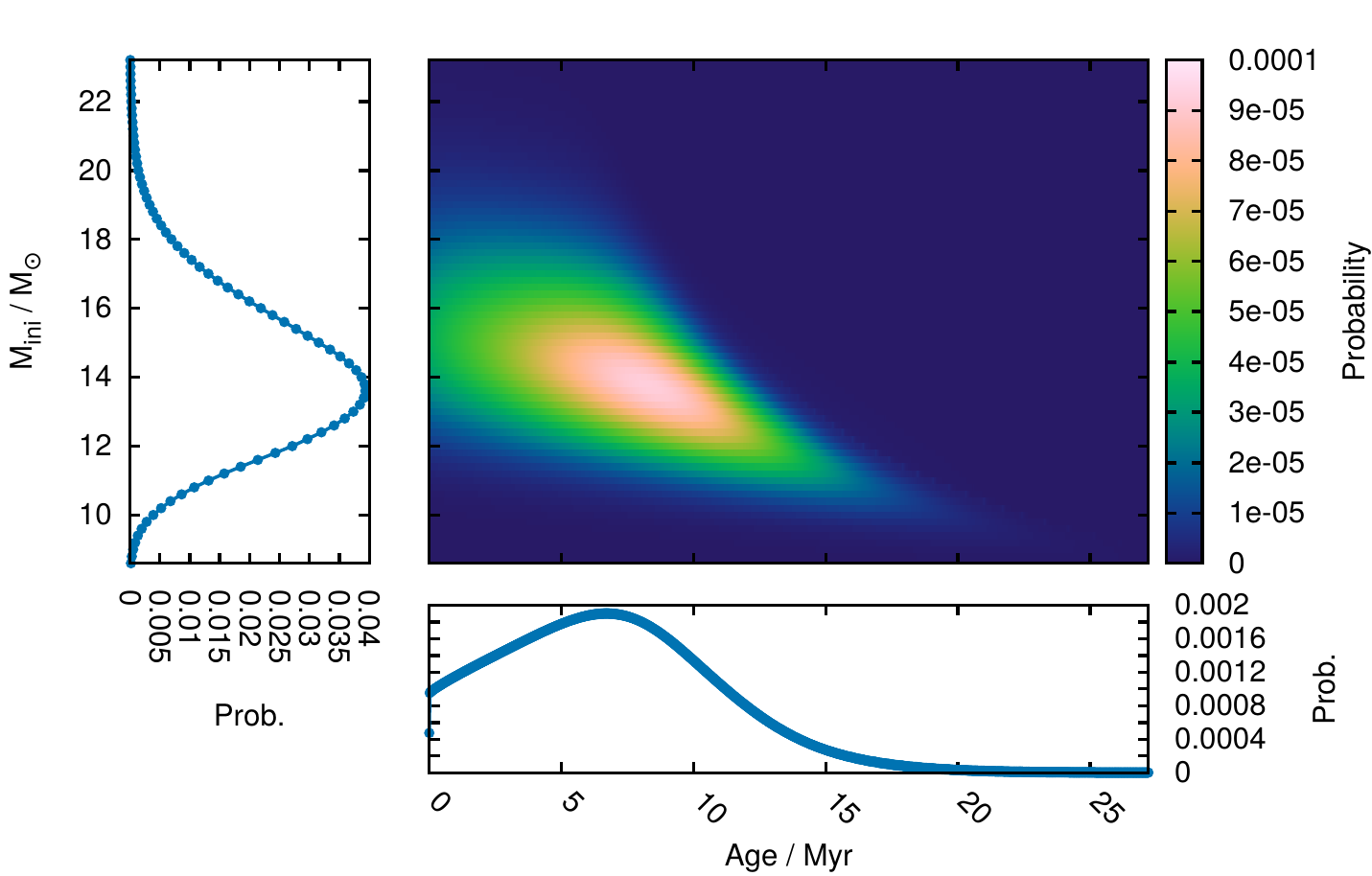}
\includegraphics[trim=0 0 0 0,clip,width=89mm]{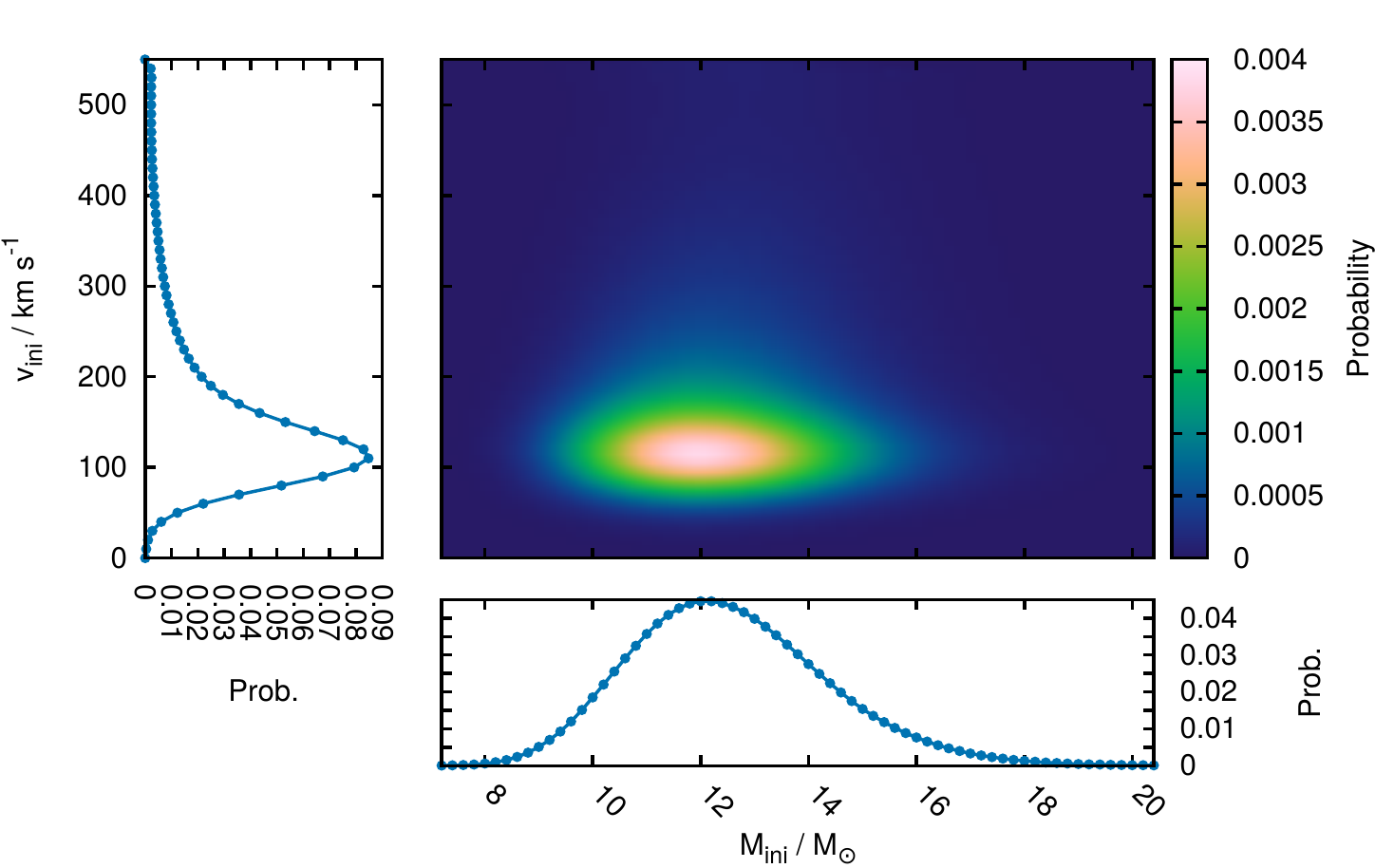}
\includegraphics[trim=0 0 0 0,clip,width=89mm]{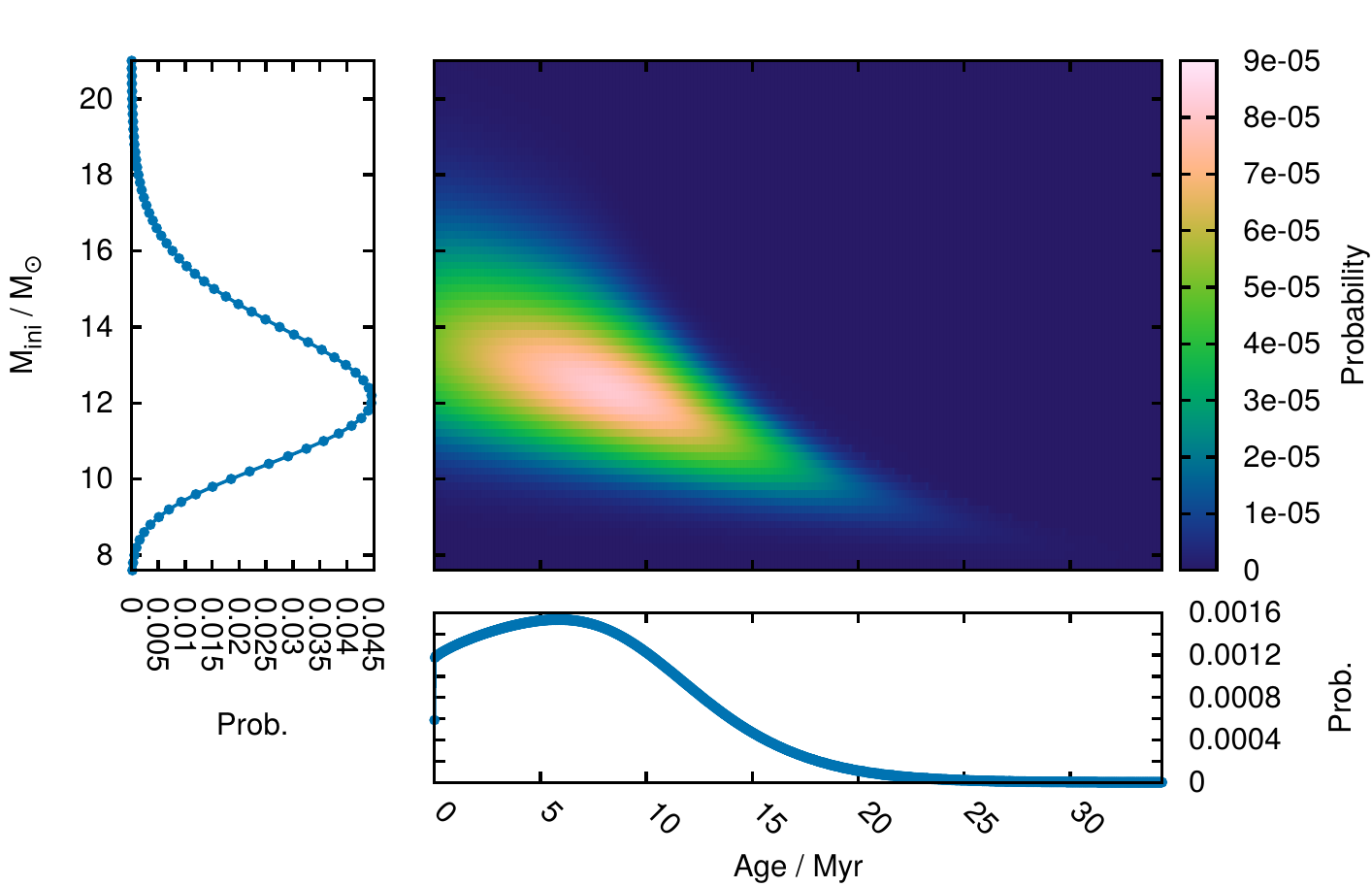}
\includegraphics[trim=0 0 0 0,clip,width=89mm]{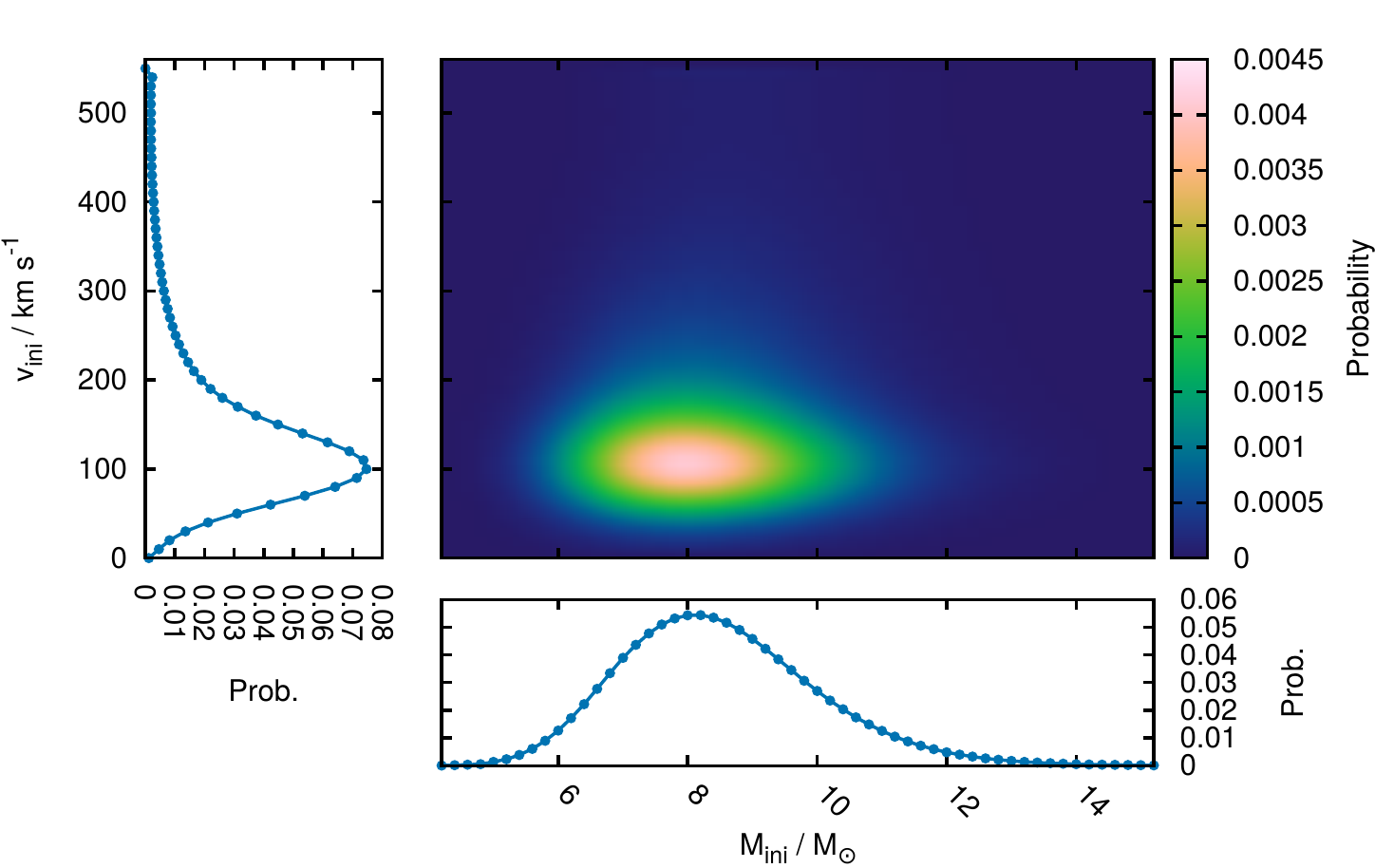}
\includegraphics[trim=0 0 0 0,clip,width=89mm]{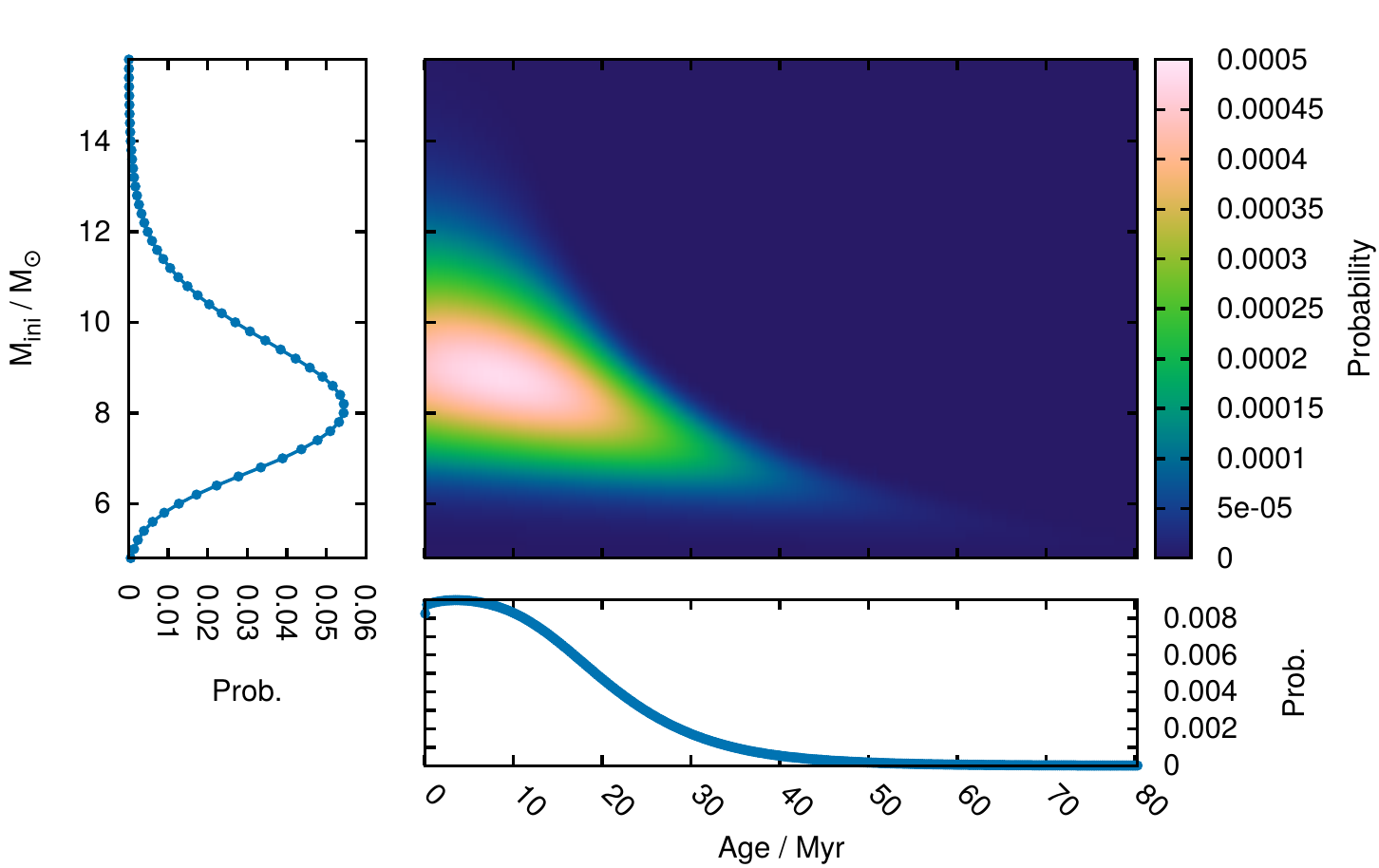}

\caption{Posterior probability maps  generated by BONNSAI for S2 (top row), S4 (middle row), S9 (bottom row).  Left column: the plot in the middle panel of each row shows the probability map in the initial mass ($M_{ini}$)--initial rotational velocity ($v_{ini}$) plane. One dimensional probability distributions for  $v_{ini}$ (left) and $M_{ini}$ (bottom) are also shown. Right column: same as the left column plot but in Age--$M_{ini}$ plane. }
\label{fig:probmap}
\end{center}
\end{figure}

%%%%%%%%%%%%%%%%%%%%%%%%%%%

\end{document}